\newif\ifpdf
\ifx\pdfoutput\undefined
\pdffalse % we are not running PDFLaTeX
\else
\pdfoutput=1 % we are running PDFLaTeX
\pdftrue \fi

\newif\iffinal
\finalfalse % Not final version
\finaltrue % Final version

\documentclass[reqno,twoside,final]{myjstatphys}
\iffinal\else\usepackage[notref,notcite]{showkeys}\fi
\usepackage{cite}
\usepackage{amsmath}
\usepackage{amsfonts}
\usepackage{amssymb}
\usepackage{graphicx}
\usepackage{graphics}
\usepackage{verbatim}

\IfFileExists{epsf.def}{\input epsf.def}{\usepackage{epsf}}

\IfFileExists{myowntimes.sty}{\usepackage{myowntimes}}
{\usepackage{times}}
\IfFileExists{mathrsfs.sty}{\usepackage{mathrsfs}}{\newcommand{\mathscr}{\mathcal}}
\DeclareFontFamily{OT1}{eufm}{} \DeclareFontShape{OT1}{eufm}{m}{n}
{<5> <6> <7> <8> <9> <10> <11> <12> <14.4> eufm10}{}
\DeclareMathAlphabet{\mathfrak}{OT1}{eufm}{m}{n}

\DeclareFontFamily{OT1}{cmss}{} \DeclareFontShape{OT1}{cmss}{m}{n}
{<5> <6> <7> <8> <9> <10> <11> <12> <13> <14.4> cmss10}{}
\DeclareMathAlphabet{\cmss}{OT1}{cmss}{m}{n}
\newenvironment{proofsect}[1]
{\par\normalfont\vskip0.3cm\noindent
{\hskip10.4mm
\sffamily\slshape#1.}}
{\qed\vspace{0.15cm}}

\theoremstyle{jsp}
\newtheorem{property}{Property}
\newtheorem{proposition}[theorem]{Proposition}

\newcommand{\textd}{\text{\rm d}\mkern0.5mu}
\newcommand{\texti}{\text{\rm i}\mkern0.7mu}
\newcommand{\texte}{\text{\rm e}\mkern0.7mu}

\newcommand{\EE}{\mathcal E}

\newcommand{\OO}{\mathcal O}

\newcommand{\RR}{\mathcal R}

\newcommand{\UU}{\mathcal U}
\newcommand{\VV}{\mathcal V}

\newcommand{\C}{\mathbb C}

\newcommand{\E}{\mathbb E}

\newcommand{\BbbH}{\mathbb H}

\newcommand{\BbbP}{\mathbb P}

\newcommand{\R}{\mathbb R}

\newcommand{\T}{\mathbb T}

\newcommand{\Z}{\mathbb Z}

\newcommand{\scrB}{\mathscr{B}}

\newcommand{\scrF}{\mathscr{F}}

\newcommand{\scrH}{\mathscr{H}}
\newcommand{\scrI}{\mathscr{I}}

\newcommand{\scrK}{\mathscr{K}}

\newcommand{\scrM}{\mathscr{M}}

\newcommand{\RE}{\text{\rm Re}\,}%{\Re \mathfrak e}\mkern2mu}
\newcommand{\IM}{\text{\rm Im}\,}%{\Im \mathfrak m}\mkern2mu}

\newcommand{\twoeqref}[2]{(\ref{#1}--\ref{#2})}

\newcommand{\bS}{{\boldsymbol S}}

\newcommand{\bm}{{\boldsymbol m}}

\newcommand{\bM}{{\boldsymbol M}}

\newcommand{\bb}{{\boldsymbol b}}
\newcommand{\bh}{{\boldsymbol h}}

\newcommand{\bw}{{\boldsymbol w}}
\newcommand{\bzero}{{\boldsymbol 0}}

\newcommand{\MF}{{\text{\rm MF}}}
\newcommand{\conv}{{\text{\rm Conv}}}

\newcommand{\hate}{{\text{\rm \^e}}}
\newcommand{\hatv}{{\text{\rm \^v}}}

\newcommand{\betaMF}{\beta_{\text{\rm MF}}}
\newcommand{\betaeff}{\beta_{\text{\rm eff}}}
\newcommand{\thetaU}{\theta_{\text{\rm U}}}
\newcommand{\thetaL}{\theta_{\text{\rm L}}}
\newcommand{\thetaI}{\theta_{\text{\rm I}}}
\newcommand{\thetaM}{\theta_{\text{\rm M}}}
\newcommand{\mstar}{m_\star}
\newcommand{\Conv}{\text{\rm Conv}\,}

\newcommand{\bz}{\boldsymbol z}

\newcommand{\betat}{\beta_{\text{\rm t}}}
\newcommand{\betac}{\beta_{\text{\rm c}}}
\newcommand{\lambdat}{\lambda_{\text{\rm t}}}
\newcommand{\bmA}{\bm_{\text{\rm A}}}
\newcommand{\bmS}{\bm_{\text{\rm S}}}
\newcommand{\eA}{e_{\text{\rm A}}}
\newcommand{\eS}{e_{\text{\rm S}}}
\newcommand{\ThetaA}{\Theta_{\text{\rm A}}}
\newcommand{\ThetaS}{\Theta_{\text{\rm S}}}
\newcommand{\betaeffA}{\betaeff^{(\text{\rm A})}}
\newcommand{\betaeffS}{\betaeff^{(\text{\rm S})}}
\newcommand{\xA}{x^{(\text{\rm A})}}
\newcommand{\xS}{x^{(\text{\rm S})}}

\newcommand{\hc}{h_{\text{\rm c}}}

\newcommand{\cc}{{\text{\rm c}}} 

\def\myffrac#1#2 in #3{\raise 2.6pt\hbox{$#3 #1$}\mkern-1.5mu\raise 0.8pt\hbox{$#3/$}\mkern-1.1mu\lower 1.5pt\hbox{$#3 #2$}}
\newcommand{\ffrac}[2]{\mathchoice%
	{\myffrac{#1}{#2} in \scriptstyle}
	{\myffrac{#1}{#2} in \scriptstyle}
	{\myffrac{#1}{#2} in \scriptscriptstyle}
	{\myffrac{#1}{#2} in \scriptscriptstyle}
}

\begin{document}

\title[Mean-field driven first-order phase transitions]
{Mean-field driven first-order phase transitions\\*[1mm]in systems with long-range interactions}

\author{Marek Biskup,\footnotemark[1] Lincoln Chayes,\footnotemark[1] and Nicholas~Crawford\footnotemark[1]}

\renewcommand{\thefootnote}{}
\footnotetext{\hglue-1.9em$\copyright$\,2004 by M.~Biskup, L.~Chayes, N.~Crawford. Reproduction, by any means, of the entire article for non-commercial purposes is permitted without charge.}
\renewcommand{\thefootnote}{\arabic{footnote}}

\footnotetext[1]{Department of Mathematics, UCLA, Los Angeles, CA 90095-1555, USA.}

\runningauthor{Biskup, Chayes and Crawford}

\begin{abstract}
We consider a class of spin systems on~$\Z^d$ with vector valued spins~$(\bS_x)$ that interact via the pair-potentials~$J_{x,y}\,\bS_x\cdot\bS_y$. The interactions are generally spread-out in the sense that the~$J_{x,y}$'s exhibit either exponential or power-law fall-off. Under the technical condition of reflection positivity and for sufficiently spread out interactions, we prove that the model exhibits a first-order phase transition whenever the associated mean-field theory signals such a transition. As a consequence, e.g., in dimensions $d\ge3$, we can finally provide examples of the 3-state Potts model with spread-out, exponentially decaying interactions, which undergoes a first-order phase transition as the temperature varies. Similar transitions are established in dimensions~$d=1,2$ for power-law decaying interactions and in high dimensions for next-nearest neighbor couplings. In addition, we also investigate the limit of infinitely spread-out interactions. Specifically, we show that once the mean-field theory is in a unique ``state,'' then in any sequence of translation-invariant Gibbs states various observables converge to their mean-field values and the states themselves converge to a product measure.
\end{abstract}


%\begin{comment}
%\medskip
\vglue-1cm
\setcounter{tocdepth}{3}
\footnotesize
\begin{list}{}
{\setlength{\topsep}{0in}\setlength{\leftmargin}{0.34in}\setlength{\rightmargin}{0.5in}}
\item[]
\tableofcontents
\end{list}
\vspace{-1.2cm}
\begin{list}{}
{\setlength{\topsep}{0in}\setlength{\leftmargin}{0.5in}\setlength{\rightmargin}{0.5in}}
\item[]
\end{list}
\normalsize
\bigskip
%\end{comment}

\smallskip

\section{Introduction}
\label{sec1}
\subsection{Motivation}
\label{sec1.1}\noindent
The understanding of the quantitative aspects of phase transitions is one of the basic problems encountered in physical (and other) sciences. Most of the existing mathematical approaches are based on the use of contour expansions via Pirogov-Sinai theory~\cite{PSa,PSb,Z} and/or the use of correlation inequalities~\cite{FFS,Simon-Phi2,Simon-vol2}. Notwithstanding, many ``practical'' scientists still rely on the so-called \emph{mean-field theory} which, in its systematic form, goes back to the work of Landau. From the perspective of mathematical physics, it is therefore desirable to shed as much light as possible on various mean-field theories and, in particular, attempt to place the subject on an entirely rigorous basis.

In a recent paper~\cite{BC}, two of us have established a direct connection between temperature-driven first-order phase transitions in certain ferromagnetic nearest-neighbor spin systems on~$\Z^d$ and their mean-field counterparts. The principal result of Ref.~\cite{BC} states that, once the mean-field theory signals a first-order phase transition, the actual system has a similar transition provided the dimension~$d$ is sufficiently large and/or the mean-field transition is sufficiently strong. Moreover, the transition happens for the values of parameters that are appropriately ``near'' the  mean-field transitional values; indeed, the various error terms tend to zero as~$d\to\infty$.

The principal goal of the present paper is two-fold.  First, we will considerably extend the scope  of systems to which the ideas of Ref.~\cite{BC} apply; i.e., we will prove discontinuous phase transitions in systems which heretofore have been beyond the reach of rigorous methods.  Second, we will in a general way expound on the \emph{mean-field philosophy}.  In particular, we will demonstrate that mean-field theory provides an asymptotic description of a certain class of systems regardless of the nature of their transitions.

Our approach is somewhat akin to the bulk of work on the so-called \emph{Kac limit} of lattice~\cite{Cassandro-Presutti,Bovier-Zahradnik1,Bovier-Zahradnik2,CFMP} as well as continuum~\cite{Kac-at-al,Lebowitz-Penrose,Lebowitz-Mazel-Presutti} systems. Here one considers finite-range interactions of unit total strength which are smeared out over a region of scale~$\ffrac1\gamma$. As~$\gamma$ tends to zero, each individual site interacts with larger and larger number of other sites and so, for~$\gamma\ll1$, one is in the position to prove that the characteristics of an actual system (e.g., the magnetization) are close to those of the corresponding mean-field theory. 
In particular, all ``approximations'' (i.e., upper and lower bounds) become exact as~$\gamma\downarrow0$.

Notwithstanding, the similarity between the Kac limit and our approach ends with the above statements: Our technique involves tight bounds on the fluctuations of the effective field while the analyses of Refs.~\cite{Cassandro-Presutti,Bovier-Zahradnik1,Bovier-Zahradnik2,CFMP} are based on coarse-graining arguments. As a consequence, we have no difficulty treating models with complicated single-spin spaces---even those exhibiting continuous internal symmetries or leading to power-law decay of correlations---or nearest-neighbor systems in large dimensions. Of course, there is a price to pay: Our technique requires the infrared bound on two-point correlation function which is presently available only for models obeying the condition of reflection positivity. Moreover, unless we assume power-law decaying interactions, the use of infrared bounds does not permit any statements in~$d=2$, while the Kac-limit approach works equally well in all~$d\ge2$.

\subsection{Models of interest}
\label{sec1.2}\noindent
For the duration of the paper, as in Ref.~\cite{BC}, we will focus on spin models with two body interactions as described by the formal Hamiltonian
\begin{equation}
\label{1.1}
\beta\scrH=-\beta\sum_{\langle x,y\rangle}J_{x,y}\,(\bS_x,\bS_y)-\sum_x(\bh,\bS_x).
\end{equation}
The various objects on the right-hand side are as follows:~$\beta$ is the inverse temperature, $\langle x,y\rangle$ denotes an unordered pair of distinct sites,~$J_{x,y}$ ($=J_{y,x}$) is the coupling constant associated with this pair, the spins~$\bS_x$ take values in a compact set~$\Omega\subset\R^n$, the (reduced) external field~$\bh$ is a vector from~$\R^n$ and $(\cdot,\cdot)$ denotes some inner product in $\R^n$. Implicit in the notation is an underlying \emph{a priori}  measure on~$\Omega$ which represents the behavior of the spins in the absence of interactions.  (In principle, the term which describes the coupling to the external field, namely the $(\bh,\bS_x)$'s, could be absorbed into the definition of the \emph{a priori} measure.  However, for \ae{}sthetic reasons, here we will often retain these terms as part of the interaction.)  

Mean-field behavior is typically anticipated in situations where fluctuations are insignificant and, on general grounds, one expects this to be the case in high dimensions.  These were precisely the operating conditions of Ref.~\cite{BC} (as well as of Refs.~\cite{BKLS,KS}) where, in a mathematically precise sense, the stipulation concerning the fluctuations was vindicated.  However, an alternative route for ramping down fluctuations is to consider ``spread out'' interactions, i.e., $J_{x,y}$'s which do not go to zero too quickly. As alluded to earlier, this alternative is, in fact, the common starting point for modern mathematical studies of phase transitions based on mean-field theory, e.g., Refs.~\cite{Cassandro-Presutti,Bovier-Zahradnik1,Bovier-Zahradnik2,Lebowitz-Mazel-Presutti,CFMP} and Refs.~\cite{Remco-Frank-Gord,Remco-Gordon,Remco-Gord2,Hara-Remco-Slade2,Sakai}.

Unfortunately, we do not have complete flexibility as to how we can spread out our interactions.  Indeed, our principal error estimate requires that the~$(J_{x,y})$ satisfy the condition of \emph{reflection positivity} (RP). Notwithstanding, the following three classes of interactions are available to our methods:
\settowidth{\leftmargini}{(1111a)}
\begin{enumerate}
\item[(1)]
\emph{Nearest along with next-nearest neighbor couplings}, i.e., potentials such that $J_{x,y}=\lambda$ if $x$ and~$y$ are nearest neighbors,~$J_{x,y}=\kappa$ with~$\lambda\ge2(d-1)|\kappa|$ if~$x$ and~$y$ are next-nearest neighbors and~$J_{x,y}=0$ in the remaining cases.
\item[(2)]
\emph{Yukawa-type potentials} of the form
\begin{equation}
\label{1.2a}
J_{x,y}=\texte^{-\mu\vert x-y\vert_1},
\end{equation}
where~$\mu>0$ and $\vert x-y\vert_1$ is the $\ell^1$-distance between~$x$ and~$y$.
\item[(3)]
\emph{Power-law decaying interactions} of the specific form
\begin{equation}
\label{1.3a}
J_{x,y}=\frac1{\vert x-y\vert_1^s},
\end{equation}
with~$s > 0$.
\end{enumerate}
Aside from these ``pure'' interactions, reflection positivity holds for
\begin{enumerate}
\item[(4)]
any combination of the above with positive coefficients.
\end{enumerate}
The derivation of the reflection-positivity property for these interactions goes back to the classic references on the subject~\cite{FSS,FILS1,FILS2}; for reader's convenience we will provide additional details in Sect.~\ref{sec3.1} and~Sect.~\ref{sec4} (Remark~\ref{rem4.4}).

We note that for all positive values of~$s$ the interactions listed in item~(3) are indeed, in the technical sense, reflection positive.  However, some values of~$s$ are not viable and others are not particularly useful.  Specifically, if $s \leq d$, then the interaction is attractive and non-summable so there is no  thermodynamics.  Thus we may as well assume that $s > d$.  Furthermore, if $d = 1$ and $s \geq 2$ or $d=2$ and $s \geq 4$ then our methods break down. With some reason:  In the one dimensional cases with $s > 2$, the results of Refs.~\cite{Mermin-Wagner,Thouless,Dyson1,Dyson2,ACCN} indicate (and in specific cases prove) that no magnetic ordering is possible.  Similarly, in the above mentioned two-dimensional cases, magnetic ordering is precluded in many systems.  

To summarize, we will impose the following limitations on our power-law interactions in Eq.~\eqref{1.3a}:
\begin{enumerate}
\item[(a)]
$s<2$ in~$d=1$,
\item[(b)]
$s<4$ in~$d=2$,
\item[(c)]
$s>d$ in all~$d\ge1$.
\end{enumerate}
Although case~(1) does not give us any real options for spreading the interaction beyond the previous recourse of taking $d\gg1$, cases~(2) and~(3) offer us the possibility to do so on a \emph{fixed} lattice. This is essentially obvious in case~(2)---just take the parameter~$\mu$ small.  As for case~(3) it is seen, after a little thought, that taking~$s$ close to~$d$ presents an additional and powerful method for smearing interactions.

\subsection{Outline of results}
\label{sec1.3}\noindent
Given the ability to smear interactions on a fixed lattice, much of the technology developed in Ref.~\cite{BC} can be applied \emph{without} the stipulation of ``$d$  sufficiently large.''  Thus it will prove possible to make statements about specific models on reasonable lattices with (more or less) reasonable interactions.  

One such ``specific'' model will be the $q$-state Potts model (see Sect.~\ref{sec2.2}).  Here, for example, we will establish a discontinuous transition between the ordered and disordered states of a $3$-state Potts model on~$\Z^3$ with interactions decaying to zero exponentially.  (And similarly for any other $q$-state Potts model on~$\Z^d$ with~$q\ge 3$ and~$d\ge 3$.) Analogous first-order phase transitions are also proved in dimensions one and two  provided we have power-law decay of the couplings as discussed above. For example, in~$d=1$, for any power-law decay exponent~$s\in(1,2)$, we produce couplings such that the~$3$-state Potts model has a first-order transition as the overall strength of the coupling varies.

As another illustration, we consider the low temperature behavior of the Blume-Capel model. The system will be described precisely in Sect.~\ref{sec3.4}, for now it suffices to say that the spins take values in $\{-1,0,+1\}$ with \emph{a priori} equal weights.  The zero temperature phase diagram of this model has a triple point where the three states of constant spin are degenerate in energy, however, as demonstrated in Ref.~\cite{Slawny}, this degeneracy is broken at finite temperatures in favor of the state dominated by the zeros.  The previous analyses of this phenomenon required rather detailed contour estimates; here we will establish similar results by relatively painless methods.

The techniques at our disposal will allow us to put to rest some small controversies which, in recent years, have been topics of some discussion.  For instance, a conjecture has been made~\cite{Kerimov96,Kerimov98} which boils down to the statement that in any one-dimensional finite-state spin system with arbitrary translation-invariant, summable interaction, the set of phase-coexistence points at positive temperatures is a \emph{subset} of the corresponding set at zero temperature. We will rule this out by our analysis of the Potts models in an external field.   

In addition to predicting first-order transitions, our mean-field framework provides an explicit description of general lattice spin systems in the limit when the interactions become highly diffuse. In particular we show that, whenever the mean-field theory is in a unique ``state,'' the magnetization and the energy density of the actual system converge to their mean-field counterparts. Moreover, \emph{every} translation invariant Gibbs state converges to a product (i.i.d.) measure with individual-spin distribution self-consistently adjusted to produce the correct value of the magnetization. (This vindicates the assumptions typically used to ``justify'' mean-field theory; see Sect.~\ref{sec2.1}.) Results in this direction have appeared before; cf Refs.~\cite{BKLS,KS}, but the main difference is that here we are \emph{not} forcing $d \to \infty$ and hence it is possible to envision a limiting system towards which we are heading.

\subsection{Organization}
\noindent
The organization of the remainder of this paper is as follows: In Sect.~\ref{sec2.1} we describe, in succinct terms, some general aspects of mean-field theory.  In Sect.~\ref{sec2.2} we discuss the mean-field theory for the Potts model in an external field---which is the primary model studied in this work. Precise results concerning these situations are the subject of Sect.~\ref{sec2.3}.

Sect.~\ref{sec3} is devoted to the statements of our main result. Specifically, in Sect.~\ref{sec3.1} we formulate a general theorem (Theorem~\ref{thm2.2}) that allows us to prove first-order phase transitions in actual lattice models with interaction~\eqref{1.1}---and RP couplings---by comparison to the associated mean-field theory. Sect.~\ref{sec3.2} provides conditions under which the mean-field theory is obtained as a limit of lattice systems when the interaction becomes infinitely spread out. Sects.~\ref{sec3.3} and~\ref{sec3.4} contain precise statements of our theorems concerning the behavior of the specific systems we study:  The zero-field $q$-state Potts models with $q\geq 3$, the same model (with~$q\ge4$) in an external field which enhances or supresses---depending on the sign---one of the states, and the Blume-Capel model near its zero-temperature triple point. Sect.~\ref{sec3.5} mentions some recent conjectures that can be addressed using our results.

The principal subject of Sect.~\ref{sec4} is to give the proof of our general results (Theorems~\ref{thm2.2} and~\ref{T-philosophy}). As part of the proof, we will discuss certain interesting convexity bounds (Sect.~\ref{sec4.1}), reflection positivity (Sect.~\ref{sec4.2}) and infrared bounds (Sect.~\ref{sec4.3}). In Sect.~\ref{sec4.5} we show how the specific interactions listed in Sect.~\ref{sec1.2} fit into our general scheme. 
Sect.~\ref{sec5} is devoted to the mathematical details of the mean-field theories for all the above mentioned models; in particular the proofs of all claims made in Sect.~\ref{sec2.3}. Sect.~\ref{sec6} then assembles all ingredients into the proofs for actual lattice systems. 

\section{Mean-field theory and the Potts model}
\noindent
Here we shall recall to mind a formalism underlying (our version of) mean-field theory and provide heuristic discussion of the basic facts. The specifics will be demonstrated on an example of the $q$-state Potts model in an external field; first somewhat informally in Sect.~\ref{sec2.2} and then precisely in Sect.~\ref{sec2.3}.

\subsection{Mean-field heuristic}
\label{sec2.1}\noindent
We will focus on the situations described by the Hamiltonian in Eq.~\eqref{1.1}.   Of course the real models must be carefully defined on $\Z^d$ as limits of finite volume measures corresponding to this Hamiltonian at inverse temperature $\beta$ \emph{and} some sort of boundary conditions.  We shall assume the reader is familiar with this basic theory
(enough of the relevant formalism can be found in Sect.~\ref{sec3.1}) and skip right to the consideration of an infinite-volume translation-invariant Gibbs state~$\mu_{\beta,\bh}$ corresponding to the Hamiltonian in Eq.~\eqref{1.1} and inverse temperature~$\beta$.  For convience we will assume here, as in the rest of this paper,
\begin{equation}
\label{Norm}
J_{x,x}=0,\quad\sum_{x\in\Z^d}|J_{0,x}|<\infty\quad\text{and}\quad\sum_{x\in\Z^d}J_{x,y} = 1.
\end{equation}
We will let~$\E_{\beta,\bh}$ denote the expectation with respect~$\mu_{\beta,\bh}$ and $\E_0$ expectation with respect to the \emph{a priori} (product) measure~$\mu_0$. (We will of course assume in the following that~$\mu_0$ is supported on more than one point.)

The principal idea is to study the distribution of one spin variable, e.g., the one at the origin of coordinates.  Let~$\bm$ denote the expected value of this spin, $\bm=\E_{\beta,\bh}(\bS_0)$. Then, conditioning on the configuration in the complement of the origin, we get the identity
\begin{equation}
\label{MFE}
\bm=\E_{\beta,\bh}\left(\frac{\E_0(\bS\, \texte^{(\bS,\beta\bm_0+\bh)})}{\E_0(\texte^{\beta(\bS,\beta\bm_0+\bh)})}\right),
\end{equation}
where $\bm_0$ is the \emph{random} variable given by the weighted average
\begin{equation}
\label{bigM}
\bm_0=\sum_{x\in\Z^d}J_{0,x}\,\bS_x.
\end{equation}
We emphasize that the expectation~$\E_{\beta,\bh}$ ``acts'' only on~$\bm_0$ while~$\E_0$ ``acts'' only on the auxiliary spin variable~$\bS$.

When all is said and done, the underlying \emph{assumption} behind the standard mean-field theories boils down to the statement that  the quantity $\bm_0$ is non-random, and therefore equal to~$\bm$. Postponing, momentarily, any discussion that concerns the validity of such an assumption, the immediate relevance is that in Eq.~\eqref{MFE} we can replace~$\bm_0$ by~$\bm$ which in turn makes the outer expectation on the right-hand side redundant.  We thus arrive at the self-consistency constraint
\begin{equation}
\label{1.6}
\bm=\frac{\E_0(\bS\, \texte^{(\bS,\beta\bm+\bh)})}{\E_0(\texte^{\beta(\bS,\beta\bm+\bh)})}
\end{equation}
which is the \emph{mean-field equation} for the magnetization.
Clearly, if it can be established that the fluctuations of $\bm_0$ are negligible, then the actual magnetization must be near a solution of Eq.~\eqref{1.6}.

In this light, our results are not that hard to understand:  In most instances where the mean-field theory predicts a discontinuous transition this prediction is showcased by the fact that Eq.~\eqref{1.6} simply does not admit continuous solutions.  Thus if
the error caused in the approximation~$\bm_0\approx \bm$ is much smaller than the discontinuities predicted in the mean-field approximation, jumps of the physical magnetization cannot be avoided.

As all of the above is predicated on the near constancy of the random variable~$\bm_0$, let us turn to a discussion of the fluctuations of this quantity.  An easy calculation shows that
\begin{equation}
\label{V}
\text{Var}(\bm_0)  =
\sum_{x,y}J_{0,x}J_{0,y}\E_{\beta,\bh}
\bigl((\bS_x,\bS_y)-|\bm|^2\bigr)
\end{equation}
where~$|\bm|^2=(\bm,\bm)$.
The quantity $\E_{\beta,\bh}((\bS_x,\bS_y)-|\bm|^2)$ is the thermal two-point correlation function which, on general grounds, may be presumed to tend to zero at large separations.  It would thus seem that the stipulation of a ``spread out interaction'' along with \emph{any} sort of decay estimate on the two-point correlations would allow us to conclude that the variance of $\bm_0$ is indeed small.  However, while explanations of this sort are satisfactory at a heuristic level, a second glance at Eq.~\eqref{V} indicates that the task is not necessarily trivial.  Indeed, of actual interest is the decay of correlations within the effective range of the interaction, which is guaranteed to be delicate.  At the core of this paper is the use of \emph{reflection positivity} to provide these sorts of estimates.

\smallskip
In many cases,  Eq.~\eqref{1.6} on its own is insufficient for understanding the behavior of a system---even at the level of mean-field theory.  Specifically, in the case of a discontinuous transition, Eq.~\eqref{1.6} will typically have multiple solutions the overall structure of which does not allow for a continuous solution.  While this may have the advantage of signaling the existence of discontinuities, it does not provide any insight as to where the discontinuities actually occur.  Thus, whenever there are multiple solutions to Eq.~\eqref{1.6}, a supplementary ``rule'' is needed to determine which of these solutions ought to be selected.

The supplement---or starting point of the whole theory depending on one's perspective---is the introduction of the \emph{mean-field free-energy function}~$\Phi_{\beta,\bh}(\bm)$ defined as follows: Let~$S(\bm)$ be the \emph{entropy function} associated with the \emph{a priori} measure on the spins. Formally, this quantity is defined by means of the Legendre transform
\begin{equation}
\label{Entropy}
S(\bm)=\inf_{\bb\in\R^n}\bigl\{G(\bb)-(\bb,\bm)\bigr\}
\end{equation}
of the cumulant generating function
\begin{equation}
\label{Gfunction}
G(\bb)=\log\E_0\bigl(\texte^{(\bb,\bS)}\bigr).
\end{equation}
The mean-field free-energy function is then defined as the difference of the energy function, $E(\bm)=-\frac {\beta}2|\bm|^2-(\bh,\bm)$, and the entropy~$S(\bm)$:
\begin{equation}
\label{PHI}
\Phi_{\beta,\bh}(\bm)=-\frac{\beta}2|\bm|^2-(\bh,\bm)-S(\bm).
\end{equation}
Then, as is not hard to see, the mean-field equation is implied by the condition that $\Phi_{\beta,\bh}$ be minimized.  Indeed, writing ~$\nabla\Phi_{\beta,\bh}(\bm)=\boldsymbol 0$ some straightforward manipulations give us
\begin{equation}
\bm=\nabla G(\beta\bm+\bh),
\end{equation}
which is exactly Eq.~\eqref{1.6}. 

Eq.~\eqref{PHI} along with the stipulation to minimize adds a whole new dimension to the theory that was defined by Eq.~\eqref{1.6}.  Foremost, in the case of multiple solutions, we now have a ``rule'' for the selection of the relevant solutions.  Beyond this, we have a framework resembling a full thermodynamical theory: A free energy---defined by evaluating $\Phi_{\beta,\bh}$ at the minimizing $\bm$---along with an entropy and energy which are the corresponding functions evaluated at this magnetization.  In fact, a secondary goal of this work is to demonstrate that this ``more complete'' mean-field theory provides an asymptotic description of the actual theories with spread out interactions.

\begin{remark}
We conclude this subsection with the remark that the mean-field theory for any particular Hamiltonian of the form \eqref{1.1} can be produced in an actual spin-system by considering the model on the \emph{complete graph}.  Explicitly, for a system with $N$ sites, we take $J_{x,y} = \frac 1N$, compute all quantities according to the standard rules of statistical mechanics and then take $N\to\infty$.   The result of this procedure is the mean-field theory described in this subsection for the thermodynamics and a limiting distribution for the spins which is i.i.d.  The connection between mean-field theory and complete graph models is well known and has been proved in numerous special cases (see, e.g., Ref.~\cite{Ellis-at-al} for a recent study of ensemble equivalence for the Potts model on the complete graph).  A complete proof for the general form of $\scrH$ given in Eq.~\eqref{1.1} appears e.g. in Sect.~5 of Ref.~\cite{BC}.
\end{remark}

\subsection{Potts models in external field}
\label{sec2.2}\noindent
The best example of a system which exhibits a rich spectrum of behaviors while remaining tractable is the Potts model in an external field. The Potts model is typically defined using discrete spin
variables $\sigma_x\in\{1,\dots,q\}$ with no apparent internal geometry.  The energy of a configuration is given by the (formal) Hamiltonian
\begin{equation}
\label{Potts}
\beta H = \beta\sum_{x,y}J_{x,y}
\delta_{\sigma_x,\sigma_y} -\sum_x h\delta_{1,\sigma_y}.
\end{equation}
Here~$\beta$ is the inverse temperature, the $J_{x,y}$'s are the coupling
constants for the system, and $\delta_{\sigma_x,\sigma_y}$ is the Kronecker delta. The reduced external field~$h$ is related to the physical external field~$\tilde h$ via $\tilde h=h/\beta$.
We have chosen only the state ``$1$'' as the state affected by the external field even though more general versions are also possible~\cite{BCK,BBCK00,BBCK1,BBCK2}.

This system is cast in the form of Eq.~\eqref{1.1} by using the \emph{tetrahedral representation}: 
We take spin variables $\bS_x\in\{\hatv_1, \dots \hatv_q\}$, where the
$\hatv_k$'s are the vertices of a unit tetrahedron in~$\R^{q-1}$. Inner products (defined the usual way for vectors in $\R^{q-1}$) between the $\hatv_k$'s satisfy
\begin{equation}
(\hatv_k,\hatv_l)=
\begin{cases}
1, \qquad &\text{if }\, k=l,
\\
\frac{-1}{q-1}, \qquad &\text{otherwise},
\end{cases}
\end{equation}
and so
\begin{equation}
\label{delta}
\delta_{\sigma_x,\sigma_y}-\frac1q=\frac{q-1}q(\bS_x,\bS_y).
\end{equation}
After similar consideration of the magnetic field terms, it is seen that the Hamiltonian in Eq.~\eqref{Potts} is manifestly of the form in Eq.~\eqref{1.1}.  To stay in accord with the classic references on the subject, e.g., Ref.~\cite{Wu}, we will keep the $q$-dependent prefactor suggested by Eq.~\eqref{delta}.  So, our official Hamiltonian for the Potts model will read
\begin{equation}
%label{ }
\beta\scrH=-\frac{q-1}{q}\beta\sum_{(x,y)}J_{x,y}\,(\bS_x,\bS_y)
-\frac{q-1}{q}h\sum_x(\hatv_1,\bS_x)
\end{equation}
with the $J$'s obeying Eq.~\eqref{Norm} and $h\in \mathbb{ \R}$.

The mean-field theory is best expressed in terms of the vector magnetization given by
\begin{equation}
\label{1.14a}
\bm = x_1 \hatv_1 + \dots + x_q\hatv_q,
\end{equation}
and the mean-field free-energy function is~\cite{Wu,BC}
\begin{equation}
\label{1.15a}
\Phi_{\beta,h}^{(q)}(\bm)= \sum_{k=1}^q\Bigl(-\frac \beta2x_k^2+x_k\log x_k\Bigr) -h x_1.
\end{equation}
Here the ``barycentric'' coordinates~$x_k$ are components of a probability vectors, i.e., we have~$x_k\ge0$ and~$x_1+\dots+x_q=1$. In the context of the Potts model on a complete graph,~$x_k$ represents the fraction of sites in the $k$-th spin state.

Let us start with a recapitulation of the zero-field case where the resulting theory is quite well known. For each~$q$ there is a number $\betaMF^{(q)}$ such that if $\beta <\betaMF^{(q)}$, the unique global minimizer is the ``most symmetric state,'' $\bm=\bzero$, while for $\beta> \betaMF^{(q)}$, there are exactly $q$~(asymmetric) global minima which are permutations of one probability vector of the form $x_1 > x_2 = \dots =x_q$. Thus we may express all quantities in terms of a \emph{scalar} magnetization, e.g., $x_1 = \frac1q + m$ and $x_k = \frac1q- \frac{m}{q-1}$, $k=2,\dots,q$. Then, when~$\beta>\betaMF^{(q)}$, the mean-field magnetization is given by~$m_\MF(\beta)=\frac{q-1}q \theta $, where~$\theta $ is the maximal positive solution to the equation
\begin{equation}
\label{1.16a}
\theta =  \frac{\texte^{\beta\theta} -1}{\texte^{\beta\theta}+q-1}.
\end{equation}
The crucial point---which can be gleaned form a perturbative analysis of Eq.~\eqref{1.16a}---is the division \emph{at}~$q=2$ of two types of behavior. In particular,~$m_\MF(\beta)$ tends to a strictly positive value as~$\beta\downarrow\betaMF^{(q)}$ for~$q>2$, while for~$q=2$ the limit value is zero. (Indeed, for~$q=2$, there \emph{are no} nontrivial solutions to Eq.~\eqref{1.16a} at~$\beta=\betaMF^{(2)} = 2$.)

\begin{remark}
\label{rem2.1}
Interestingly, the values of $\betaMF^{(q)}$ and the limit value $m_\MF(\betaMF^{(q)})$ are explicitly computable:
\begin{equation}
\label{eq1}
\betaMF^{(q)}= 2\,\frac{q-1}{q-2}\log(q-1), \qquad m_\MF(\betaMF^{(q)}) = \frac{q-2}q.
\end{equation}
This observation goes back to at least Ref.~\cite{Wu}.
\end{remark}

Let us now anticipate, without going to details, what happens for $h\ne0$. (The full-blown statements and proofs will appear in Sect.~\ref{sec2.3} and Sect.~\ref{sec5}, respectively.) We will capitalize on the principle that local minimizers are stable to small changes in parameters. Consider~$q\ge3$ and~$h\ne0$ such that~$|h|\ll1$. The overall situation cannot differ too drastically from the zero-field case; the only distinction is that for~$h>0$ only one of the ``$h=0$ asymmetric minimizers'' is allowed while for~$h<0$ the same minimizer is suppressed in favor of the remaining~$q-1$ ones. On the other hand, for~$h$ positive and large, it is clear that the minimizer of~$\Phi_{\beta,h}^{(q)}(\bm)$ will be unique no matter what~$\beta$ is. Thus, for~$h>0$ we should have a line of mean-field first-order phase transitions which terminates at a finite value of~$h$. On general grounds, the terminal point is expected to be a critical point.

Next, let us consider~$h<0$ with~$|h|\gg1$. The situation at~$h=-\infty$ is clear; this is just the ($q-1$)-state Potts model. Thus for finite but large~$|h|$, we can see a clear distinction between $q=3$ and~$q>3$. In the former cases, the mean-field transition should be Ising like and hence continuous. In the latter case, the transition should be discontinuous. Thus, the~$q=3$ line should break at a \emph{tricritical} point followed by a line of continuous transitions while for~$q>3$ there will be an unbroken line of discontinuous mean-field phase transitions. 

Aside from general interest, the key motivation for obtaining such detailed knowledge about~$m_\MF$ is as follows: Under specific conditions on \eqref{1.1}, virtually all that has just been discussed pertaining to discontinuous transitions in these systems can be established with rigor in the spread out  ``real'' systems. (On the downside is the fact that virtually nothing pertaining to the continuous transition can be proved by these methods.)
To illustrate let us consider the transition at~$h>0$ when~$q$ is large.
The mean-field picture is as follows: A non-convexity of~$\Phi_{\beta,h}^{(q)}(\bm)$ develops when~$\beta$ is of order unity, but it does not ``touch down'' until~$\beta$ is appreciable (of order~$\log q$). However, the existence of a non-convexity suggests that a strong-enough magnetic field can tilt the balance in favor of a magnetized state, even for~$\beta$'s of order unity. This is indeed the case for the MFT as our detailed calculations later show. As a consequence of the general techniques presented here, this result from the MFT will be processed into a theorem for actual systems.

\subsection{Precise statements for mean-field Potts model}
\label{sec2.3}\noindent
Our precise results for the mean-field theory of the Potts model in an external field are summarized into two theorems; one for positive fields and the other for negative fields.

\begin{theorem}[Positive fields]
\label{TH1}
Let~$q\ge3$, let~$\bm$ and the probability vector~$(x_1,\dots,x_q)$ be related as in Eq.~\eqref{1.14a} and let $\Phi_{\beta,h}^{(q)}(\bm)$ denote the function from Eq.~\eqref{1.15a}.
Let $\hc$ denote the quantity
\begin{equation}
\label{hc}
\hc = \log q-\frac{2(q-2)}{q}.
\end{equation}
Then there is a continuous function $\beta_+^{(q)}\colon(0,\hc)\to(0,\infty)$ such that
\settowidth{\leftmargini}{(1111)}
\begin{enumerate}
\item[(1)]
For all $(\beta,h)$ such that either $h \geq \hc$ or $\beta \neq \beta_+^{(q)}(h)$,
there is a unique global minimizer of $\Phi_{\beta,h}^{(q)}(\bm)$ with $x_2 = \dots = x_q$.  The quantity~$x_1$ corresponding to this minimizer is strictly larger than the mutual value of the~$x_k$'s for~$k=2,\dots,q$.
\item[(2)]
For all $h < \hc$, there are two distinct global minimizers of $\Phi_{\beta,h}^{(q)}(\bm)$ at $(\beta_+^{(q)}(h),h)$.
\item[(3)]
For $(\beta,h)$ such that $h \geq \hc$ or $\beta \neq \beta_+^{(q)}(h)$,  let $x_1=x_1(\beta,h)$ denote the first coordinate of the global minimizer of  $\Phi_{\beta,h}^{(q)}(\bm)$.  Then $(\beta,h) \mapsto  x_1(\beta,h)$ is continuous with well-defined but distinct (one-sided) limits at~$(\beta,h)=(\beta_+^{(q)}(h),h)$. 
Furthermore, writing $x_1=\frac1q + m$, the quantity $\theta=\frac q{q-1}m$ obeys the equation
\begin{equation}
\label{ET1.1}
\theta =  \frac{\texte^{\beta\theta+h} -1}{\texte^{\beta \theta +h} +q-1}.
\end{equation}
in the region of uniqueness.  At the points $(\beta_+^{(q)}(h),h)$, both limiting values obey this equation.
\item[(4)]
The function $h\mapsto\beta_+^{(q)}(h)$ is strictly decreasing on $(0,\hc)$ with limit values $\beta_+^{(q)}(h)\uparrow \betaMF^{(q)} =2\frac{q-1}{q-2}\log(q-1)$ as $h\downarrow 0$ and $\beta_+^{(q)}(h) \downarrow \frac{4(q-1)}{q}$ as $h\uparrow \hc$.
\end{enumerate}
\end{theorem}

In order to preserve uniformity of exposition, we will restrict the statement of negative-field results to $q\ge4$.

\begin{theorem}[Negative fields]
\label{TH2} 
Let~$q\ge4$, let~$\bm$ and the probability vector~$(x_1,\dots,x_q)$ be related as in Eq.~\eqref{1.14a} and let $\Phi_{\beta,h}^{(q)}(\bm)$ denote the function from Eq.~\eqref{1.15a}.
Then we have:
\settowidth{\leftmargini}{(1111)}
\begin{enumerate}
\item[(1)] \label{it2.1}
All global minima are permutations in the last $q-1$ variables of vectors with the representation
\begin{equation}
\label{ET2.1} 
x_1 < x_{2} = \dots = x_{q-1} \leq x_{q}.
\end{equation}
\end{enumerate}
\smallskip
Moreover, there exists a function $\beta_-^{(q)}\colon(-\infty,0)\to(0,\infty)$ such that
the following hold:
\smallskip
\settowidth{\leftmargini}{(1111)}
\begin{enumerate}
\item[(2)] \label{it2.2}(Symmetric Minimum) For all $\beta <\beta_-^{(q)}(h)$, there is a unique global minimum and it has $x_2=\dots=x_q$. Moreover, if~$m$ is such that $x_1 = \frac1q-m$ and $x_{k} = \frac1{q} + \frac m{q-1}$, for all $k=2,\dots,q$, then $\theta=\frac q{q-1}m$
corresponds to a global minimum when
\begin{equation}
\label{E2.1} 
\theta =  \frac{\texte^{\beta\theta-h}-1}{(q-1)\texte^{\beta\theta-h} +1}.
\end{equation}
There is only one $\theta\in [0,\frac1{q-1}]$ for which Eq.~\eqref{E2.1} holds.
\item[(3)] 
\label{it2.3} (Asymmetric Minima) For all $\beta>\beta_-^{(q)}(h)$,
we have $q-1$ global minima.  These are permutations in the last
$q-1$ variables of a single minimum whose coordinate
representation takes the form
\begin{equation}
\label{E2.21}
x_1 < x_{2} = \dots = x_{q-1} < x_{q}.
\end{equation}
\item[(4)] At~$\beta=\beta_-^{(q)}(h)$ there are~$q$ global minima. 
One of these is of the type described in~(2)---namely,
the symmetric minimum---while the other~$q-1$ are of the type
described in~(3). 
\item[(5)] The function $h\mapsto\beta_-^{(q)}(h)$ is
strictly increasing and continuous.  Moreover, we have the limits
\begin{equation}
%\label{}
\lim_{h\to-\infty}\beta_-^{(q)}(h) = \betaMF^{(q-1)}
\quad\text{and}\quad
\lim_{h\uparrow 0}\beta_-^{(q)}(h) = \betaMF^{(q)}
\end{equation}
\end{enumerate}
\end{theorem}

Theorem~\ref{TH1} is proved in Sect.~\ref{sec5.3} and Theorem~\ref{TH2} is proved in Sect.~\ref{sec5.4}. The corresponding statement for the actual lattice systems is the subject of  Theorem~\ref{T2}.

\section{Main results}
\label{sec3}\noindent
Here we give the statements of the principal theorems which apply to any model whose interaction is of the type \eqref{1.1}. Then we apply these to the Potts and Blume-Capel models.

\subsection{General theory}
\label{sec3.1}\noindent
We begin by a precise definition of the class of models we consider. Let~$\Omega$ be a compact subset of~$\R^n$, with the inner product denoted by~$(\cdot,\cdot)$, and let~$\Conv(\Omega)$ denote the convex hull of~$\Omega$. Let~$\mu_0$ be a Borel probability measure on~$(\Omega,\scrB)$ that describes the \emph{a priori} distribution of the individual spins.
We will consider spin configurations~$(\bS_x)$ from~$\Omega^{\Z^d}$ and, abusing the notation slightly, use~$\mu_0$ to denote also the corresponding \emph{a priori} product measure. 

To define the interacting spin system, let us pick a finite set~$\Lambda\subset\Z^d$, a spin configuration~$\bS_\Lambda\in\Omega^\Lambda$ in~$\Lambda$ and the ``boundary condition'' $\bS_{\Lambda^\cc}\in\Omega^{\Lambda^\cc}$. For each~$\bh\in\R^n$ and each~$\beta>0$, we then define the finite-volume Hamiltonian $\scrH_\Lambda(\bS_\Lambda,\bS_{\Lambda^\cc})$ by
\begin{equation}
\label{1.1fin}
\beta\scrH_\Lambda(\bS_\Lambda,\bS_{\Lambda^\cc})=-\beta\sum_{\begin{subarray}{c}
\langle x,y\rangle\\ x\in\Lambda,y\in\Z^d
\end{subarray}}
J_{x,y}\,(\bS_x,\bS_y)
-\sum_{x\in\Lambda}(\bh,\bS_x).
\end{equation}
The first sum goes over all unordered pairs of distinct sites~$\langle x,y\rangle$ at least one of which is contained in~$\Lambda$.

The above Hamiltonian can now be used to define the finite-volume Gibbs measure~$\nu_\Lambda^{(\bS_{\Lambda^\cc})}$ on spin configuration from~$\Omega^\Lambda$ by
\begin{equation}
%\label{}
\nu_\Lambda^{(\bS_{\Lambda^\cc})}(\textd \bS_\Lambda)
=\frac{\texte^{-\beta\scrH_\Lambda(\bS_\Lambda,\bS_{\Lambda^\cc})}}
{Z_\Lambda^{(\bS_{\Lambda^\cc})}(\beta,\bh)}\mu_0(\textd \bS_\Lambda),
\end{equation}
where the normalizing constant $Z_\Lambda^{(\bS_{\Lambda^\cc})}(\beta,\bh)$ is the partition function. Of particular interest are the (weak subsequential) limits of these measures as~$\Lambda$ expands to fill out the entire~$\Z^d$. These measures obey the DLR-conditions~\cite{Georgii} and are generally referred to as (infinite-volume) Gibbs measures. In this formalism, \emph{phase coexistence} is said to occur for parameters~$\beta$ and~$\bh$ if there is more than one limiting Gibbs measure. Under these conditions the system is said to exhibit a \emph{first-order phase transition}.

\smallskip
We proceed by formulating the precise conditions under which our results will be proved. To facilitate our next definition, for each lattice direction $\ell\in\{1,\dots,d\}$, let~$\BbbH_\ell$ denote the half-space
\begin{equation}
%\label{}
\BbbH_\ell=\{x=(x_1,\dots,x_d)\in\Z^d,\,x_\ell>0\}.
\end{equation}
We will use~$\vartheta^{(\ell)}$ to denote the reflection~$\vartheta^{(\ell)}\colon\BbbH_\ell\to\Z^d\setminus\BbbH_\ell$ defined explicitly by the formula
$\vartheta^{(\ell)}(x_1,\dots,x_d)=(x_1,\dots,x_{\ell-1},1-x_\ell,x_{\ell+1},\dots,x_d)$.

\begin{definition}[RP ``through bonds'']
\label{def2}
Consider a collection of coupling constants $(J_{x,y})_{x,y\in\Z^d}$. We say that these are RP if the following conditions hold:
\settowidth{\leftmargini}{(11a111)}
\begin{enumerate}
\item[(1)]
(translation invariance) for any~$x,y\in\Z^d$ we have $J_{x,y}=J_{0,y-x}$.
\end{enumerate}
Moreover, for any lattice direction $\ell\in\{1,\dots,d\}$,
\begin{enumerate}
\item[(2)]
(reflection invariance) for any~$x,y\in\BbbH_\ell$ we have
\begin{equation}
%\label{}
J_{x,y}=J_{\vartheta^{(\ell)}x,\vartheta^{(\ell)}y}.
\end{equation}
\item[(3)]
(reflection positivity) if~$f\colon\BbbH_\ell\to\R$ is absolutely summable with
\begin{equation}
%\label{}
\sum_{x\in\BbbH_\ell}f(x)=0,
\end{equation}
then
\begin{equation}
\label{2.2}
\sum_{\begin{subarray}{c}
x\in\BbbH_\ell\\y\in\Z^d\smallsetminus\BbbH_\ell
\end{subarray}}
J_{x,y}f(x)f(\vartheta^{(\ell)}y)\ge0.
\end{equation}
\end{enumerate}
\end{definition}

Given a translation-invariant Gibbs measure, we use the word \emph{magnetization} to denote the expectation of the spin at the origin. The statement of our general result can then be viewed as a restriction on the possible values of the magnetization. However, not all magnetizations that can be physically produced are (provably) accessible to our methods. The reason is that the underlying Gibbs states for which our techniques work will have to satisfy the conditions of reflection positivity---in particular, they have to be obtained as weak limits of torus states. Our next item of business will be to define precisely the set of ``allowed values'' of the magnetization.

We will proceed as in Ref.~\cite{BC}. Let~$Z_\Lambda(\beta,\bh)$ be the partition function in volume~$\Lambda$---the boundary condition is irrelevant---and let~$F(\beta,\bh)$ denote the (physical) free energy defined as the limit of~$-\frac1{|\Lambda|}\log Z_\Lambda$ as~$\Lambda$ increases to fill the entire~$\Z^d$ (in the sense of van Hove~\cite{Georgii}). The function~$F(\beta,\bh)$ is jointly concave, so we may let~$\scrK_\star(\beta,\bh)$ denote the set of all pairs~$[e_\star,\bm_\star]$ such that
\begin{equation}
%\label{}
F(\beta+\triangle\beta,\bh+\triangle\bh)-F(\beta,\bh)\le e_\star\,\triangle\beta+(\bm_\star,\triangle\bh)
\end{equation}
for any~$\triangle\beta\in\R$ and any~$\triangle\bh\in\R^n$. Now~$\scrK_\star(\beta,\bh)$ is a convex set so we let~$\scrM_\star(\beta,\bh)$ to denote the set of values~$\bm_\star$ for which there exists an~$e_\star$ such that~$[e_\star,\bm_\star]$ is an extreme value of~$\scrK_\star(\beta,\bh)$. Our main theorem then reads:

\begin{theorem}
\label{thm2.2}
Consider the spin system on~$\Z^d$ with the Hamiltonian \eqref{1.1} such that the couplings~$(J_{x,y})$ are~RP, the inverse temperature~$\beta>0$ and external field~$\bh\in\R^n$. For each~$k\in[-\pi,\pi]^d$, let~$\hat J(k)=\sum_{x\in\Z^d}J_{0,x}\texte^{\texti k\cdot x}$ and recall that~$\hat J(0)=1$ by Eq.~\eqref{Norm}. Then for any~$\bm_\star\in\scrM_\star(\beta,\bh)$,
\begin{equation}
\label{2.4}
\Phi_{\beta,\bh}(\bm_\star)\le\inf_{\bm\in\Conv(\Omega)}\Phi_{\beta,\bh}(\bm)+\beta n\frac\kappa2\scrI,
\end{equation}
where~$n$ is the (underlying) dimension of the spin-space,~$\kappa=\max_{\bS\in\Omega}|\bS|^2$ and
\begin{equation}
\label{2.5}
\scrI=\int_{[-\pi,\pi]^d}\frac{\textd k}{(2\pi)^d}
\frac{|\hat J(k)|^2}{1-\hat J(k)}.
\end{equation}
\end{theorem}

\smallskip
The useful aspect of Theorem~\ref{thm2.2} is that the error term~$\EE=\beta n\frac\kappa2\scrI$ can be made small by appropriate adjustment of parameters. A general statement of this sort appears in Proposition~\ref{prop4.7} but, typically, these conditions have to be verified on a case by case basis. Let us tend to the details of these adjustments later and, for the time being, simply assume that~$\EE$ is small. Then, along with the obvious supplement of Eq.~\eqref{2.4}, $\Phi_{\beta,\bh}(\bm_\star)\ge\inf_{\bm\in\Conv(\Omega)}\Phi_{\beta,\bh}(\bm)$, we have learned that the allowed values of the magnetization in the \emph{physical} system nearly minimize the \emph{mean-field} free energy. In this sense, the mean-field theory already provides a quantitatively accurate description of the physical system once~$\EE\ll1$. In Sects.~\ref{sec3.3}-\ref{sec3.4} we will use this fact to prove a first-order phase transitions in a few models of interest.

To demonstrate the use of Theorem~\ref{thm2.2}, let us consider the ``evolution'' of a typical MFT phase transition, in which two local minima of~$\Phi_{\beta,\bh}$ exchange roles of the global minimizer as~$\beta$ varies.
Specifically, let~$\bmS(\beta)$ and~$\bmA(\beta)$ be local minima of~$\Phi_{\beta,\bh}$---one of which is always global---for~$\beta$ near some~$\betat$, and suppose that~$\Phi_{\beta,\bh}(\bmA)>\Phi_{\beta,\bh}(\bmS)$ for~$\beta>\betat$ and \emph{vice versa} for~$\beta<\betat$. Then Theorem~\ref{thm2.2} can be applied under the condition that, outside some small neighborhoods of~$\bmS(\beta)$ and~$\bmA(\beta)$ for~$\beta\approx\betat$, no magnetizations have a free energy within~$\EE$ of the absolute minimum. For~$\beta\gtrapprox\betat$, this stipulation applies even to the neighborhood of~$\bmS(\beta)$ and, for~$\beta\lessapprox\betat$, to the neighborhood of~$\bmA(\beta)$. Then, Theorem~\ref{thm2.2} tells us that in the region~$\beta\lessapprox\betat$, the actual magnetization is near~$\bmS(\beta)$, for~$\beta\approx\betat$ it could be near~$\bmS$ or~$\bmA$, and for~$\beta\gtrapprox\betat$ it is only near~$\bmA(\beta)$.
On general grounds, as long as the difference~$\bmA-\bmS$ is bounded uniformly away from zero, somewhere near~$\betat$ there has to be a point of phase coexistence.

\subsection{Mean-field philosophy}
\label{sec3.2}\noindent
In this section we will state some general facts about spin systems and their mean-field analogues. The stipulations that govern this section are rather mild; first we will assume that the Hamiltonian is of the form \eqref{1.1} with the $J_{x,y}$'s satisfying the conditions of reflection positivity. Second, we will assume that the associated mean-field free-energy function defined in Eq.~\eqref{PHI} has a unique minimizer. Finally, we will investigate the small-$\scrI$ behavior of these models. The preferred viewpoint is a fixed dimension~$d$ with parameters~$\mu$---as defined in Eq.~\eqref{1.2a}---tending to zero or~$s$---as defined in Eq.~\eqref{1.3a}---tending to~$d$.

We note that special cases (usually restricted to concrete models) have been addressed elsewhere; see, in particular, Ref.~\cite{KS} and references therein, but there the only mechanism to force~$\scrI\to0$ was the~$d\to\infty$ limit which we find \ae{}sthetically somewhat unsatisfactory. Another possibility is to consider the aforementioned Kac limit which more or less boils down to infinite smearing out of the interaction. A contour-based analysis of this limit has been carried out, but the technical aspects have so far been overcome only for very specific models~\cite{Cassandro-Presutti,Bovier-Zahradnik1,Bovier-Zahradnik2,Lebowitz-Mazel-Presutti,CFMP}. Here we provide a general result in this direction under the sole condition of reflection positivity.

\begin{theorem}[Mean-field philosophy]
\label{T-philosophy}
Consider the spin system as described above and let~$\Phi_{\beta,\bh}$ be as in Eq.~\eqref{PHI}. Suppose that the parameters~$\beta>0$ and~$\bh\in\R^n$ are such that~$\Phi_{\beta,\bh}$ has a unique minimizer~$\bm$ on~$\Conv(\Omega)$ in Eq.~\eqref{PHI}. Let~$(J_{x,y}^{(n)})$ be a sequence of coupling constants that are~RP and obey Eq.~\eqref{Norm}, and let~$\langle-\rangle_{\beta,\bh}^{(n)}$ be a sequence of translation and rotation-invariant Gibbs states corresponding to these couplings. If the sequence of integrals~$\scrI_n$, obtained from~$(J_{x,y}^{(n)})$ via Eq.~\eqref{2.5}, satisfies
\begin{equation}
%\label{}
\scrI_n\to0\quad\text{as}\quad n\to\infty,
\end{equation}
then we have the following facts:
\settowidth{\leftmargini}{(1111)}
\begin{enumerate}
\item[(1)]
The actual magnetization tends to~$\bm$, i.e.,
\begin{equation}
\label{ideol1}
\langle\bS_0\rangle_{\beta,\bh}^{(n)}\,\underset{n\to\infty}\longrightarrow\,\bm.
\end{equation}
\item[(2)]
The energy density tends to its mean-field value, i.e.,
\begin{equation}
\label{ideol2}
\bigl\langle(\bS_0,\tfrac\beta2\bm_0+\bh)\bigr\rangle_{\beta,\bh}^{(n)}
\,\underset{n\to\infty}\longrightarrow\,E(\bm),
\end{equation}
where~$\bm_0$ is as in Eq.~\eqref{bigM} and~$E(\bm)$ is as in Sect.~\ref{sec2.1}.
\end{enumerate}
In particular, in the limit~$n\to\infty$, the spin variables at distinct sites become independent with distribution given by the product of the titled measures
\begin{equation}
\label{ideol4}
\texte^{(\bS,\beta\bm+\bh)-G(\beta\bm+\bh)}\mu_0(\textd\bS).
\end{equation}
Here~$\mu_0$ is the \emph{a priori} measure.
\end{theorem}

\smallskip
The preceding---as is the case in much of the principal results of this paper---reduces (the $\scrI\to0$ limit of) the full problem to a detailed study of the associated mean-field theory. Two specific models will be analyzed in great detail shortly (see Sects.~\ref{sec3.3} and~\ref{sec3.4}); let us mention two other well known (or well studied) examples.

First are the $O(n)$ spin systems at zero external field. Here each~$\bS_x$ takes values on the unit sphere in~$\R^n$ with \emph{a priori} uniform measure. In the mean-field theory of these models, the scalar magnetization $m(\beta)$ vanishes for~$\beta$ less than some~$\betac$ while for~$\beta\ge\betac$ it is the maximal positive solution of a certain transcendental equation (see, e.g., Ref.~\cite{KS}). In particular, this solution rises continuously from zero according to
\begin{equation}
%\label{}
|m(\beta)|=(\beta-\betac)^{\ffrac12}\,\bigl[C(n)+o(1)\bigr],\qquad \beta\downarrow\betac.
\end{equation}
By Theorem~\ref{T-philosophy}, the actual magnetization converges to this function but, unfortunately, our control is not strong enough to rule out the possibility of small discontinuities (which vanish as~$\scrI\to0$).

A less well known but very interesting example is the \emph{cubic model} where the spins point to the center of a face on an~$r$-dimensional unit hypercube, i.e.,~$\bS_x\in\Omega=\{\pm\hate_1,\dots,\pm\hate_r\}$. For~$r>3$ the transition in this model is first order (and was analyzed in Ref.~\cite{BC}). The case~$r=2$ reduces to an Ising system but the borderline case,~$r=3$, while still continuous, features a somewhat anomalous (namely, tricritical) behavior. Indeed, for this system, the mean-field magnetization obeys
\begin{equation}
%\label{}
|m(\beta)|=(\beta-\betac)^{\ffrac14}\,\bigl[C+o(1)\bigr],\qquad \beta\downarrow\betac,
\end{equation}
where~$\betac=3$. Once again, the actual magnetization converges to such a function but the control is not sufficient to rule out small discontinuities.

While these sorts of results do not establish \emph{any} critical behavior in particular systems, they could represent a first step in proving that a variety of (mean-field) critical behaviors are possible.

\subsection{Results for the Potts model}
\label{sec3.3}\noindent
Our first result concerns the zero-field $q$-state Potts model with~$q\ge3$.
Let~$F(\beta,h)$ denote the free energy of the Potts model with the Hamiltonian in Eq.~\eqref{Potts} and let $\mstar(\beta)$ be the quantity 
\begin{equation}
\label{Potts-magn}
m_\star(\beta) = \frac{\partial}{\partial h^+} F(\beta,h)\Bigl|_{h=0}-\frac1q.
\end{equation}
(An alternative definition of~$m_\star(\beta)$ would be the limiting probability that the spin at the origin is ``1'' in the state generated by the boundary spins all set to ``1.'')
Let~$m_\MF=m_\MF(\beta)$ be related to the maximal positive solution~$\theta$ of Eq.~\eqref{1.16a} by $m_\MF=\frac{q-1}q\theta$. Then we have:

\begin{theorem} 
\label{T1}
Let~$q\ge3$ be fixed. For each~$\epsilon>0$ there exists~$\delta>0$ with the following property: For any $d\ge1$ and any collection of coupling constants $(J_{x,y})$ on~$\Z^d$ that are~RP, obey~\eqref{Norm} and for which the integral~$\scrI$ in Eq.~\eqref{2.5} satisfies~$\scrI\le\delta$, there exists a number~$\betat\in(0,\infty)$ such that 
\begin{equation}
\label{eq3.17}
|\betat- \betaMF^{(q)}| \leq \epsilon
\end{equation}
holds and such that the physical magnetization~$\mstar=\mstar(\beta)$ of the corresponding $q$-state Potts model obeys the bounds
\begin{equation}
m_\star(\beta) \leq \epsilon\quad \text{for} \quad \beta < \betat
\end{equation}
and
\begin{equation}
\label{eq3.19}
|m_\star(\beta)-m_\MF(\beta)| \leq \epsilon\quad \text{for} \quad
\beta > \betat.
\end{equation}
In particular, whenever the integral~$\scrI$ is sufficiently small, $\beta\mapsto m_\star(\beta)$
undergoes a jump near the value~$\betaMF^{(q)}$. A similar jump occurs (at the same point) in the energy density.
\end{theorem}

This statement extends Theorem~2.1 of Ref.~\cite{BC} to a class of spread-out RP interactions. (A minor technical innovation is that the bound in Eq.~\eqref{eq3.19} holds uniformly.)
As a consequence, we are finally able to provide examples of interactions for which the~$q=3$ state Potts models in dimension~$d=3$ can be proved to have a first-order transition. Similar conclusion holds for all~$q\ge3$ but, unfortunately, our requirements on the ``smallness'' of the corresponding parameters are not uniform in~$q$. 

In~$d=1$, we show that the long-range Potts models with power-law decaying interactions go first order once the exponent of the power-decay is between one and two. Models in this category have been studied in Ref.~\cite{Newman-Schulman} in the context of percolation; the domination techniques of, e.g., Ref.~\cite{ACCN} then imply the existence of a low temperature phase. However, the percolation-based approach alone is unable to tell whether the transition is discontinuous or not. Some additional discussion is provided in Sect.~\ref{sec3.5}.

\smallskip
Our next item of interest will be the same system in an external field, as described by the full Hamiltonian \eqref{Potts}. For reasons alluded to in Sect.~\ref{sec2.2}, we will restrict our attention to the~$q\ge4$ cases.

\begin{theorem}
\label{T2}
Let~$q\ge4$ be fixed and let us consider the~$q$-state Potts model with coupling constants~$J_{x,y}$ that are RP and obey Eq.~\eqref{Norm}. Then there exists~$\delta_0>0$ and a function~$h_0\colon(0,\delta_0]\to[0,\hc)$, where $\hc$ is as in Eq.~\eqref{hc}, such that if \eqref{2.5} obeys~$\scrI\le\delta$ with some~$\delta\le\delta_0$, then there exists a function~$\betat\colon(-\infty,h_0)\to(0,\infty)$ with the following properties:
\settowidth{\leftmargini}{(1111)}
\begin{enumerate}
\item[(1)]
A first-order transition (accompanied by a discontinuity in the energy density and the magnetization) occurs at the parameters $(h,\betat(h))$, for any external field~$h\in(-\infty,h_0)$.
\item[(2)]
Let~$\mstar(\beta,h)$ be the ``spin-1 density'' defined by the right partial derivative~$\frac\partial{\partial h^+}F(\beta,h)$. Then there exists an~$h_1=h_1(\delta)<0$ such that $h\mapsto\mstar(\beta,h)$ has a discontinuity at field strength~$\tilde h$ such that~$\beta=\betat(\tilde h)$ provided that~$\tilde h\in(h_1,h_0)$. 
\end{enumerate}
The function~$h_0$ is decresing while~$h_1$ is increasing. Moreover, $\lim_{\delta\downarrow0}h_0(\delta)=\hc$ and~$\lim_{\delta\downarrow0}h_1(\delta)=-\infty$.
\end{theorem}

The second part of the theorem asserts that, even if state~``$1$'' is suppressed by the field, the order-disorder transition will be felt by the ``spin-1 density'' $m_\star(\beta,h)$.
There is no doubt in our mind that the restriction to~$h\ge h_1$ in this claim is only of technical nature. Our lack of control for~$h$ very large negative stems from the fact that the jump in the mean-field counterpart of~$\mstar(\beta,h)$ decreases exponentially with~$|h|$ as~$h\to-\infty$. Theorems~\ref{T1} and~\ref{T2} are proved in Sect.~\ref{sec6}.

\subsection{Results for the Blume-Capel model}
\label{sec3.4}\noindent
The Blume-Capel model is a system whose spins~$\sigma_x$ take values in the set $\Omega=\{-1,0, 1\}$ with \emph{a priori} equal weights. The Hamiltonian is given most naturally in the form
\begin{equation}
\label{BC-Ham}
\beta\mathscr{H}(\sigma) = \beta\sum_{\langle x,y\rangle}J_{x,y}(\sigma_x - \sigma_y)^2
- \lambda \sum_x(\sigma_x)^2 - h
\sum_x\sigma_x.
\end{equation}
As is easy to see, a temporary inclusion of the terms proportional to~$(\sigma_x)^2$ into the single-spin measure shows that this Hamiltonian is indeed of the general form in Eq.~\eqref{1.1}.

If we consider the situation at zero temperature ($\beta = \infty$)  with~$\lambda$ and $h$ finite we see that in the $(\lambda,h)$-plane there are three regions of constant spin which minimize~$\beta\mathscr{H}(\sigma)$.  The regions all meet at the point $h = 0$, $\lambda = 0$; tentatively we will call the origin a triple point (and the lines phase boundaries).  Ostensibly one would wish to establish that this entire picture persists at finite temperature. However, we will confine attention to the line $h = 0$ which is of the greatest interest.  We will show, both in the context of mean-field theory and, subsequently, realistic systems that there is indeed a finite temperature first order transition at some~$\lambdat(\beta)$.  Of significance is the fact that this occurs at a~$\lambdat$ which is \emph{strictly} positive; i.e., for $1 \ll \beta < \infty$, the point $\lambda = 0$ lies inside the phase which is dominated by zeros.

We remark that results of this sort are far from new; indeed the proof of this and similar results represented one of the early triumphs of low temperature techniques Ref.~\cite{Slawny}.  The physical reason behind the shifting of the phase boundary is the enhanced ability of the ``zero'' phase over the plus and minus phases to harbor elementary excitations.  Interestingly, in spite of the fact that our method relies on \emph{suppression} of fluctuations, the  corresponding entropic stabilization is nevertheless manifest in our derivation. In addition, while the contour-based approaches require a non-trivial amount of ``low temperature labor'' to ensure that the interactions between excitations are limited,  our methods effortlessly incorporate whatever interactions may be present.

\smallskip
To simplify our discussion, from now on we will focus on the situation at zero external field, i.e.,~$h=0$, and suppress~$h$ from the notation. First let us take a look at the mean-field theory. Here we find it useful to express the relevant quantities in terms of mole fractions~$x_1,x_0,x_{-1}$ of the three spin states in~$\Omega$. To within an irrelevant constant, the mean-field free-energy function is
\begin{equation}
\label{BC1.1}
\Phi_{\beta,\lambda} = 4\beta x_1 x_{-1} + \beta x_0(1-x_0)
+\lambda x_0+\sum_{\sigma=\pm1,0}x_\sigma\log x_\sigma.
\end{equation}
Here we have used the fact that $x_1 + x_0 + x_{-1} = 1$. Our main result concerning the mean-field theory of the Blume-Capel model is now as follows:

\begin{theorem}
\label{BCT1} 
For all~$\beta\ge0$ and all~$\lambda\in\R$, all local minima of~$\Phi_{\beta,\lambda}$ obey the equations
\begin{equation}
\label{3.15a}
x_1\texte^{4\beta x_{-1}}=x_{-1}\texte^{4\beta x_1}=x_0\texte^{\beta(1-2x_0)+\lambda}.
\end{equation}
Moreover, there exists a~$\beta_0<\infty$ such that for all~$\beta\ge\beta_0$, any such (local) minimum is of the form that two components of~$(x_1,x_0,x_{-1})$ are very near zero and the remaining one is
near one. Explicitly, there exists a constant~$C<\infty$ such that
\begin{enumerate}
\item[(1)]
If~$x_0$ is the dominant index, then~$x_1=x_{-1}=\frac12(1-x_0)$ and we have that~$(1-x_0)\le C \texte^{-\beta+\lambda}$.
\item[(2)]
If~$x_1$ is the dominant index, then~$x_{-1}\le C\texte^{-4\beta}$ while~$x_0\le C\texte^{-\beta-\lambda}$. A corresponding statement is true for the situation when~$x_{-1}$ is dominant.
\end{enumerate}
Furthermore, consider two local minima at~$(\beta,\lambda)$, one dominated by~$x_0$ and the other dominated by~$x_1$. Let~$\phi_0(\beta,\lambda)$ be the mean-field free energy corresponding to the former minimum and let~$\phi_1(\beta,\lambda)$ be that corresponding to the latter minimum. Then
\begin{equation}
\label{3.16b}
\phi_0(\beta,\lambda)-\phi_1(\beta,\lambda)=\lambda-\texte^{-\beta+\lambda}+O(\beta\texte^{-2\beta})
\end{equation}
where $O(\beta\texte^{-2\beta})$ denotes a quantity bounded by a constant times~$\beta\texte^{-2\beta}$ for all~$\lambda$ in a neighborhood of the origin. 
In particular, for all~$\beta$ sufficiently large there exists $\lambda_\MF(\beta)=\texte^{-\beta}+O(\beta\texte^{-2\beta})$ such that the global minimizes of~$\Phi_{\beta,\lambda}$ have~$x_{\pm1}\ll1$ for~$\lambda<\lambda_\MF(\beta)$ and~$x_0\ll1$ for $\lambda>\lambda_\MF(\beta)$.
\end{theorem}

Theorem~\ref{BCT1} is proved in Sect.~\ref{sec5.1}. Next we will draw our basic conclusions about the actual system:

\begin{theorem}
\label{BCT2}
Consider the Blume-Capel model in Eq.~~\eqref{BC-Ham}, with zero field ($h=0$), inverse temperature~$\beta$ and the coupling constants~$(J_{x,y})$ that are RP and obey Eq.~\eqref{Norm}. Let~$\scrI$ be the integral in Eq.~\eqref{2.5}. There exist constants~$\beta_0\in(0,\infty)$ and~$C<\infty$ such that if~$\beta\ge\beta_0$ and~$\beta\scrI\ll\texte^{-\beta}$, then there is a function~$\lambdat\colon[\beta_1,\beta_2]\to\R$ satisfying~$|\lambdat(\beta)-\texte^{-\beta}|<\beta\scrI$ such that any translation-invariant Gibbs state $\langle-\rangle_{\beta,\lambda}$ obeys
\begin{enumerate}
\item[(1)]
$\langle\sigma_x^2\rangle_{\beta,\lambda}\le C\texte^{-\beta}$ if~$\lambda<\lambdat(\beta)$,
\item[(2)]
$\langle\sigma_x^2\rangle_{\beta,\lambda}\ge 1-C\texte^{-\beta}$ if~$\lambda>\lambdat(\beta)$.
\end{enumerate}
Moreover, at~$\lambda=\lambdat(\beta)$, there exist three distinct, translation-invariant Gibbs states $\langle-\rangle_{\beta,\lambda}^\sigma$, with $\sigma\in\{+1,0,-1\}$, the typical configuration of which contains fraction at least~$1-C\texte^{-\beta}$ of the corresponding spin state.
\end{theorem}

We remark that the phase transition happens at a value of~$\lambda$ which (at least for~$\beta\gg1$) is strictly positive. This demonstrates the phenomenon of entropic suppression (of~$\pm1$ ground states at~$\lambda=0$) established previously in Ref.~\cite{Slawny} by the contour-expansion techniques. The entropic nature of the above transition is also manifested by the fact that the free-energy ``gap'' separating the distinct states \emph{decreases} as~$\beta\to\infty$. This is the reason why, to maintain uniform level of control, we need~$\scrI$ to be smaller for smaller temperatures. Theorem~\ref{BCT2} is proved in Sect.~\ref{sec6}.

\subsection{Discussion}
\label{sec3.5}\noindent
We close this section with a discussion of some conjectures that can be addressed via the above theorems.

Starting with the intriguing results in Ref.~\cite{Kerimov93} and culminating in Refs.~\cite{Kerimov96,Kerimov98}, A.~Kerimov formulated the following conjecture (we quote verbatim from the latter pair of references): ``Any one-dimensional model with discrete (at most countable) spin space and with a unique ground state has a unique Gibbs state if the spin space of this model is finite or the potential of this model is translationally invariant.'' The conclusions of Theorem~\ref{T1} manifestly demonstrate that this conjecture fails for the~1D Potts model in external field. Indeed, for~$q\ge3$, $h>0$ and interactions decaying like~$1/r^s$ with~$s\in(1,2)$ which are RP and satisfy the condition that the integral in Eq.~\eqref{2.5} is sufficiently small, the Potts model has phase coexistence at some positive temperature. However, it is clear that this system enjoys a unique ground state.

In a recent paper~\cite{Berger}, N.~Berger considered random-cluster models with parameter~$q$ and interactions between sites~$x$ and~$y$ decaying as~$|x-y|^{-s}$, where~$d<s<2d$. He proved, among other results, that at the percolation threshold there is no infinite cluster in the measure generated by the  free boundary conditions. For ordinary percolation (i.e.,~$q=1$), this implies continuity of the infinite cluster density. As to the wired boundary conditions, for~$q=2$---i.e., the Ising model---the classic results of Refs.~\cite{AF,ABF} show that the magnetization vanishes continuously once the model is in the ``mean-field regime'' $s\in(1,\ffrac32)$. However, for general random-cluster models with~$q>1$ and wired boundary conditions, the situation remained open. 

While we cannot quite resolve the situation \emph{at} the percolation threshold, our results prove that, for sufficiently spread out random-cluster models with RP couplings, there is a point where the free and wired densities are indeed different. To resolve the full conjecture from Ref.~\cite{Berger}, one would need to establish that the only place such a discontinuity can occur is at the percolation threshold.

Our third application concerns the problem of partition function zeros of the Potts model in a \emph{complex} external field with~$\RE h<0$. Here there have been numerical results~\cite{Kim-Creswick} claiming that no such zeros occur for the nearest-neighbor 2D Potts model with~$q\le7$. On the basis of the classic Lee-Yang theory~\cite{YL,LY}, absence of such zeros would imply analyticity of the spin-1 density. The results of Refs.~\cite{BBCKK1,BBCKK2,BBCK0,BBCK1,BBCK2} rule this out for~$q$ very large and Theorem~\ref{T2}(2) also makes this impossible for reasonable values of~$q$ and sufficiently spread-out interactions (of course, for~$d=1,2$ this requires a  power-law interaction).

\section{Proofs: General theory}
\label{sec4}\noindent
The goal of this section is to prove Theorems~\ref{thm2.2} and~\ref{T-philosophy}. In Sect.~\ref{sec4.1} we present some general convexity results that provide the framework for the derivation of our results. However, the driving force of our proofs are the classic tools of reflection positivity and infrared bounds  which are reviewed (and further developed) in Sects.~\ref{sec4.2} and~\ref{sec4.3}. The principal results of this section are Theorem~\ref{Th4.1} and Lemmas~\ref{lemma4.2},~\ref{lemma-key} and~\ref{lemma4.8}.

\subsection{Convexity bounds}
\label{sec4.1}\noindent
We begin with an intermediate step to Theorem~\ref{thm2.2} which gives an estimate on how far above the mean-field free energy evaluated at a \emph{physical} magnetization is from the absolute minimum.

\begin{theorem}
\label{Th4.1} 
Suppose $(J_{x,y})$ are translation and rotation
invariant couplings on~$\Z^d$ such that Eq.~\eqref{Norm} holds. 
Let $\nu_{\beta,\bh}$ be a translation and rotation-invariant, infinite volume Gibbs measure
corresponding to $\beta \geq 0$ and $\bh\in\R^n$.  
Let $\langle-\rangle_{\beta,\bh}$ denote the expectation with respect to~$\nu_{\beta,\bh}$ and let $\bm_\star=\langle\bS_0\rangle_{\beta,\bh}$. Then
\begin{equation}
\label{eq4.1} 
\Phi_{\beta,\bh}(\bm_\star) \leq \inf_{\bm \in
\conv(\Omega)} \Phi_{\beta,\bh}(\bm) +
\frac\beta2\big\{ \bigl\langle (\bS_0,\bm_0)\bigr\rangle_{\beta,\bh}-|\bm_\star|^2 \bigr\},
\end{equation}
where~$\bm_0=\sum_{x\in\Z^d} J_{0,x} \bS_x$.
\end{theorem}

\begin{proofsect}{Proof}
The proof is very similar to that of Theorem~$1.1$ of Ref.~\cite{BC}.
Let~$\Lambda$ be a box of~$L\times\dots\times L$ sites in~$\Z^d$ and let~$\bM_\Lambda$ be the total spin in~$\Lambda$, i.e., $\bM_\Lambda=\sum_{x\in\Lambda}\bS_x$. Let us also recall the meaning of the mean-field quantities from \twoeqref{Entropy}{PHI}. The starting point of our derivations is the formula
\begin{equation}
\label{eq4.2} 
\texte^{|\Lambda|G(\bb)} = \bigl\langle\, \texte^{(\bb, \bM_\Lambda) +
\beta \scrH_{\Lambda}(\bS_\Lambda|\bS_{\Lambda^\cc})}
Z_{\Lambda}(\bS_{\Lambda^\cc})\bigr\rangle_{\beta, \bh},
\qquad \bb\in\R^n,
\end{equation}
which is obtained by invoking the DLR conditions for the Gibbs state~$\nu_{\beta,\bh}$. Here~$\scrH_{\Lambda}(\bS_\Lambda|\bS_{\Lambda^\cc})$ is as in Eq.~\eqref{1.1fin} and~$Z_\Lambda(\bS_{\Lambda^\cc})$ is a shorthand for the partition function in~$\Lambda$ given~$\bS_{\Lambda^\cc}$.

The goal is to derive a lower bound on the right-hand side of Eq.~\eqref{eq4.2}.
First we provide a lower bound on $Z_{\Lambda}(\bS_{\Lambda^\cc})$ which is independent of boundary conditions. To this end, let $\langle-\rangle_{0,\bb}$ denote expectation with respect to the product measure
\begin{equation}
%\label{}
\texte^{(\bb,\bM_\Lambda)-|\Lambda|G(\bb)}\prod_{x\in\Lambda}\mu_0(\textd\bS_x)
\end{equation}
and let~$\bm_\bb$ denote the expectation of any spin in~$\Lambda$ with respect to this measure. Jensen's inequality then gives us
\begin{equation}
\begin{aligned}
Z_{\Lambda}(\bS_{\Lambda^\cc})&=e^{|\Lambda|G(\bb)}\bigl\langle 
\texte^{-(\bb,\bM_\Lambda)-\beta\scrH_{\Lambda}(\bS_\Lambda|\bS_{\Lambda^\cc})}\bigr\rangle_{0,\bb}
\\
&\ge
\texte^{|\Lambda|[G(\bb)-(\bb,\bm_\bb)]}
\,\texte^{-\langle\beta\scrH_{\Lambda}(\bS_\Lambda|\bS_{\Lambda^\cc})\rangle_{0,\bb}}.\end{aligned}
\end{equation}
Now, \twoeqref{Entropy}{Gfunction} imply that~$G(\bb)-(\bb,\bm_\bb)=S(\bm_\bb)$, while the absolute summability of~$x\mapsto J_{0,x}$ implies that for all $\epsilon>0$ there is a $C_1<\infty$, depending on~$\epsilon$, the $J_{x,y}$'s and the diameter of~$\Omega$, so that
\begin{equation}
\label{eq4.4}
-\bigl\langle\,\beta\scrH_{\Lambda}(\bS_\Lambda|\bS_{\Lambda^\cc})\bigr\rangle_{0,\bb}
\geq |\Lambda|E(\bm_\bb) -
\beta\epsilon |\Lambda| - \beta C_1|\partial\Lambda|,
\end{equation}
with~$E(\bm_\bb)$ denoting the mean-field energy function from Sect.~\ref{sec2.1}. (Note that we used also the normalization condition \eqref{Norm}.) Invoking Eq.~\eqref{PHI} and optimizing over all~$\bb\in\R^n$, we thus get
\begin{equation}
\label{eq4.3} 
Z_{\Lambda}(\bS_{\Lambda^\cc}) \geq  \texte^{-|\Lambda| F_\MF(\beta, \bh) - \beta\epsilon |\Lambda| -
\beta C_1|\partial \Lambda|},
\end{equation}
where~$F_\MF(\beta, \bh)$ is the absolute minimum of $\Phi_{\beta, \bh}(\bm)$ over all~$\bm\in\Conv(\Omega)$.

Having established the desired lower bound on the partition function, we now plug the result into Eq.~\eqref{eq4.2} to get
\begin{equation}
\label{4.6s}
\texte^{|\Lambda|G(\bb)} \geq \bigl\langle \texte^{(\bb, \bM_\Lambda) + \beta
\scrH_{\Lambda}(\bS_\Lambda|\bS_{\Lambda^\cc})}\bigr\rangle_{\beta, \bh}\,
\texte^{- |\Lambda| F_\MF(\beta, \bh) - \beta\epsilon |\Lambda| -\beta C_1|\partial \Lambda|}.
\end{equation}
The expectation can again be moved to the exponent using Jensen's inequality, now taken with respect to measure~$\nu_{\beta,\bh}$. 
Invoking the translation and rotation invariance of this Gibbs state, bounds similar to Eq.~\eqref{eq4.4} imply
\begin{multline}
\label{eq4.6} 
\bigl\langle\,\beta\scrH_{\Lambda}(\bS_\Lambda|\bS_{\Lambda^\cc})\bigr\rangle_{\beta,\bh}
\\
\geq -|\Lambda|\biggl(\, \sum_{x \in\Z^d}\frac\beta2 J_{0,x} \bigl\langle(\bS_x,\bS_0)\bigr\rangle_{\beta,\bh} + (\bh,\bm_\star) - \epsilon \biggr) 
- C_2|\partial \Lambda|.
\end{multline}
Plugging this back into Eq.~\eqref{4.6s}, taking logarithms, dividing by~$|\Lambda|$ and letting $|\Lambda|\to\infty$ (with~$|\partial\Lambda|/|\Lambda|\to0$) followed by~$\epsilon\downarrow0$, we arrive at the bound
\begin{equation}
G(\bb)-(\bb, \bm_\star)\geq -
\frac\beta2\sum_{x \in \Z^d} J_{0,x}
\bigl\langle(\bS_x,\bS_0) \bigr\rangle_{\beta,\bh}-
(\bh,\bm_\star) - F_\MF(\beta, \bh).
\end{equation}
Optimizing over~$\bb$ gives
\begin{equation}
S(\bm_\star) - (\bh,\bm_\star) \leq \frac\beta2\sum_{x\in\Z^d}J_{0,x} 
\bigl\langle(\bS_x,\bS_0)\bigr\rangle_{\beta,\bh} +
F_\MF(\beta, \bh)
\end{equation}
from which Eq.~\eqref{eq4.1} follows by subtracting $\frac\beta2|\bm_\star|^2$ on both sides.
\end{proofsect}

Similar convexity estimates allow us to establish also the following bounds between the energy density and fluctuations of
the weighted magnetization~$\bm_0$:

\begin{lemma}
\label{lemma4.2}
Let $\kappa=\sup_{\bS \in \Omega}(\bS, \bS)$ and let~$(J_{x,y})$ be a collection of couplings satisfying Eq.~\eqref{Norm}. For each~$\beta>0$ and~$\bh\in\R^n$ there exists a number~$\varkappa=\varkappa(\beta,\bh)$ such that for any translation and rotation invariant Gibbs state~$\langle-\rangle_{\beta,\bh}$ we have
\begin{equation}
\label{eq4.11b}
\beta \varkappa\,\bigl\langle\,|\bm_0-\bm_\star|^2\bigr\rangle_{\beta,\bh}
\le \bigl\langle(\bS_0,\bm_0)\bigr\rangle_{\beta,\bh}-|\bm_\star|^2
\le\beta\kappa\,\bigl\langle\,|\bm_0-\bm_\star|^2\bigr\rangle_{\beta,\bh},
\end{equation}
where~$\bm_0=\sum_{x\in\Z^d}J_{0,x}$ and~$\bm_\star=\langle\bS_0\rangle_{\beta,\bh}$.

\end{lemma}

\begin{proofsect}{Proof}
We begin with a rewrite of the correlation function in the middle of Eq.~\eqref{eq4.11b}. First, using the DLR equations to condition on the spins in the complement of the origin, we have
\begin{equation}
\label{eq4.24}
\bigl\langle(\bm_0,\bS_0)\bigr\rangle_{\beta,\bh}
=\bigl\langle(\bm_0,\nabla G(\beta\bm_0+\bh)\bigr\rangle_{\beta,\bh}.
\end{equation}
Next, our hypotheses imply that $\bm_\star = \langle \bm_0 \rangle_{\beta,\bh}=\langle\nabla G(\beta \bm_0 + \bh)\rangle_{\beta,\bh}$, and so
\begin{multline}
\label{eq4.25}
\bigl\langle(\bm_0,\nabla G(\beta\bm_0+\bh)\bigr\rangle_{\beta,\bh}-|\bm_\star|^2
\\
=
\bigl\langle(\bm_0-\bm_\star, \nabla G(\beta \bm_0 + \bh) -
\nabla G(\beta \bm_\star + \bh))\bigr\rangle_{\beta,\bh}.
\end{multline}
For the rest of this proof, let $\varXi$ abbreviate the inner product in the expectation on the right-hand side.

We will express~$\varXi$ using the mean value theorem
\begin{equation}
%\label{}
\varXi=\bigl(\bm_0-\bm_\star,[\nabla\nabla G(\bb)](\bm_0-\bm_\star)\bigr),
\end{equation}
where~$\bb$ is a point somewhere on the line between~$\beta\bm_0+\bh$ and~$\beta\bm_\star+\bh$.
The double gradient~$\nabla\nabla G(\bb)$ is a matrix with components $(\nabla\nabla G(\bb))_{i,j}= \langle
S_0^{(i)}S_0^{(j)}\rangle_{0,\bb}-\langle
S_0^{(i)}\rangle_{0,\bb}\langle
S_0^{(j)}\rangle_{0,\bb}$. As was shown in~Ref.~\cite{BC}, the $\ell^2$-operator
norm of $\nabla\nabla G(\bb)$ is bounded by $\kappa=\sup_{\bS \in \Omega}(\bS, \bS)$ and so we have
\begin{equation}
\label{eq4.25b}
\varXi\le\beta\kappa\,|\bm_0-\bm_\star|^2.
\end{equation}
Taking expectations on both sides, and invoking Eqs.~\twoeqref{eq4.24}{eq4.25}, this proves the upper bound in Eq.~\eqref{eq4.11b}. 

To get the lower bound we note that, $\mu_0$ almost surely, the double gradient $\nabla\nabla G(\bb)$ is positive definite on the linear subspace generated by vectors from~$\Omega$. (We are using that~$\Omega$ \emph{is} the support of the \emph{a priori} measure~$\mu_0$.) Since $\beta\bm_0+\bh$ takes values in a compact subset of this subspace, we have
\begin{equation}
%\label{}
\varXi\ge\beta\varkappa\,|\bm_0-\bm_\star|^2
\end{equation}
for some (existential) constant $\varkappa>0$. Taking expectations, the left inequality in \eqref{eq4.11b} follows.
\end{proofsect}

%%%%
\begin{comment}
A simple consequence, which we will find useful in the proof of uniformity of the bounds in Theorem~\ref{T1}, of Theorem~\ref{Th4.1} is the following correlation inequality which does not seem to be well known:

\begin{corollary}
%\label{cor}
Let~$\nu_{\beta,\bh}$ be a Gibbs state, let~$\langle-\rangle_{\beta,\bh}$ be the corresponding expectation and let us abbreviate~$\bm_\star=\langle\bS_0\rangle_{\beta,\bh}$. Under the conditions of Theorem~\ref{Th4.1}, the energy density of~$\nu_{\beta,\bh}$ exceeds~$E(\bm_\star)$, i.e.,
\begin{equation}
%\label{}
\bigl\langle(\bS_0,\bm_0)\bigr\rangle_{\beta,\bh}\ge\frac12|\bm_\star|^2.
\end{equation}
where~$\bm_0=\sum_{x\in\Z^d}J_{0,x}\bS_x$.
\end{corollary}

\begin{proofsect}{Proof}
This is an immediate consequence of Eq.~\eqref{eq4.1} and the definition of the ``weighted'' magnetization~$\bm_0$.
\end{proofsect}
\end{comment}

We emphasize that in its present form, the bounds~\eqref{eq4.1} and~\eqref{eq4.11b} 
are essentially of complete generality.  Underlying
most of the derivations in this paper is the observation that the
variance term on the right-hand side of Eq.~\eqref{eq4.11b} is
sufficiently small. Via~Eq.~\eqref{eq4.1}, the physical magnetization~$\bm_\star$ is then forced to be near
one of the near minima of the mean-field free energy. 
This reduces the problem of proving discontinuous phase transitions to:
\begin{enumerate}
\item[(1)] controlling the variance term in Eq.~\eqref{eq4.11b},
\item[(2)] a detailed analysis of the minimizers of $\Phi_{\beta,\bh}$.
\end{enumerate}
For (1), we will use the method of reflection positivity/infrared
bounds discussed in the following subsections. 
As mentioned before, this does impose some restrictions 
on our interactions and our Gibbs states.  Part~(2) is model specific and, for the Potts and Blume-Capel models, is the subject of Sect.~\ref{sec5}.

\subsection{Reflection positivity}
\label{sec4.2}\noindent 
Our use of reflection positivity (RP) will require that we temporarily restrict our model to the torus~$\T_L$ of~$L\times\dots\times
L$ sites. In order to define the interaction potential on this
torus, we recall that the~$J_{x,y}$'s are translation invariant and define their ``periodized'' version by
\begin{equation}
\label{2.1a} 
J^{(L)}_{x,y}=\sum_{z\in\Z^d}J_{x,y+Lz},
\end{equation}
where~$Lz$ is the site whose coordinates are $L$-multiples of those of~$z$.
The torus version of the Hamiltonian~\eqref{1.1} is then defined by
\begin{equation}
\label{2.1} 
\beta \scrH_L(\bS)=-\sum_{\begin{subarray}{c}
\langle x,y\rangle\\x,y\in\T_L
\end{subarray}}
\beta J^{(L)}_{x,y}(\bS_x, \bS_y) -\sum_{x\in\T_L}(\bS_x, \bh).
\end{equation}
(Here, as in Eq.~\eqref{1.1}, the first sum is over all unordered pairs of sites.)
Let~$\BbbP_L$ denote the Gibbs measure on~$\Omega^{\T_L}$ whose
Radon-Nikodym derivative with respect to the \emph{a priori} spin
distribution~$\mu_0(\textd\bS)$ is the properly
normalized~$\texte^{-\beta\scrH_L(\bS)}$.

Let us suppose that~$L$ is even and let us temporarily
regard~$\T_L$ as a periodized box~$\{1,\dots,L\}^d$. Let~$\T_L^+$
be those sites whose~$i$-th coordinate ranges between~$1$
and~$L/2$ and let~$\T_L^-$ be the remaining sites. The two parts
of the torus are related to each other by a reflection in the
``hyperplane''~$P$ that separates the two halves from each other. (The
geometrical image of the plane has two components.) Given such a plane~$P$, we
let~$\scrF_P^+$ denote the $\sigma$-algebra of events that depend
on the configuration in~$\T_L^+$, and similarly for~$\scrF_P^-$
and~$\T_L^-$. 

Let~$\vartheta_P$ denote the reflection
taking~$\T_L^+$ onto~$\T_L^-$ and \emph{vice versa} (cf.~the definition of~$\vartheta^{(k)}$ in Sect.~\ref{sec3.1}).  In the
natural way,~$\vartheta_P$ induces an operator~$\vartheta_P^\star$
on the set of real-valued functions on $(\Omega^{\T_L})$.
Then we have:

\begin{definition}[RP on torus]
\label{def1} We say that~$\BbbP_L$ is reflection positive if for
every plane~$P$ as described above and any two bounded,
$\scrF_P^+$-measurable random variables $X$ and~$Y$,
\begin{equation}
\label{4.7}
\E_L\bigl(X\vartheta_P^\star(Y)\bigr)=\E_L\bigl(Y\vartheta_P^\star(X)\bigr)
\end{equation}
and
\begin{equation}
\label{2.3} \E_L\bigl(X\vartheta_P^\star(X)\bigr)\ge 0.
\end{equation}
Here~$\E_L$ is the expectation with respect to~$\BbbP_L$.
\end{definition}

Condition~\eqref{2.3} in the above definition is often too
complicated to be verified directly. Instead we verify a
convenient sufficient condition which we will state next:

\begin{lemma}
\label{lemma2.1} Consider a collection of coupling
constants~$(J_{x,y})_{x,y\in\Z^d}$ satisfying the
properties of Definition~\ref{def2} in Sect.~\ref{sec3.1}.  Then
the measure~$\BbbP_L$, defined on~$\T_L$ using the periodized
coupling constants from Eq.~\eqref{2.1a}, is reflection positive in the
sense of Definition~\ref{def1}.
\end{lemma}

\begin{proofsect}{Proof}
This is a multidimensional version of Proposition~3.4 of~\cite{FILS1}.
\end{proofsect}

\begin{remark}
\label{rem4.4}
We note that the three classes of interactions listed in
Sect.~\ref{sec1.2} are reflection positive. For the most part, interactions of this sort were discussed in Ref.~\cite{FILS1}; however, for reader's convenience, we provide the relevant calculations below. 

\smallskip\noindent
(1) \textsl{Nearest-neighbor/next-nearest neighbor couplings}:
Consider a function~$f\colon\BbbH_1\to\C$ which is nonzero only on the sites of~$\BbbH_1$ that are adjacent to~$\Z^d\setminus\BbbH_1$. (By inspection of Eq.~\eqref{2.2}, for nearest and next-nearest neighbor interactions, this is the most general function that need to be considered.) Pick~$\eta\in\R$ and consider the function
\begin{equation}
%\label{}
g_j(x)=f(x)+\eta f(x+\hate_j),\qquad j=2,\dots,d,
\end{equation}
and define a collection of coupling constants~$(J_{x,y})$ by the formula
\begin{equation}
%\label{}
\sum_{\begin{subarray}{c}
x\in\BbbH_1\\y\in\Z^d\smallsetminus\BbbH_1
\end{subarray}}
J_{x,y}\,\overline{f(x)}f(\vartheta^{(1)}y)=\sum_{j=2,\dots,d}\,
\sum_{x\in\BbbH_1}\overline{g_j(x)}g_j(x)
\end{equation}
Now the right-hand side is clearly positive and so the~$J_{x,y}$'s satisfy the condition in Eq.~\eqref{2.2}. 

It remains to identify the explicit form of these coupling constants. Let~$x\in\BbbH_1$ be a boundary site and let~$x'=\vartheta^{(1)}x$ be its nearest neighbor in~$\Z^d\setminus\BbbH_1$. First we note that, for each~$x$ and~$j$, there is an interaction of ``strength''~$\eta$ between~$x$ and its next-nearest neighbor~$x'+\hate_j$ and a similar interaction between~$x$ and the site~$x'-\hate_j$. So, the next-nearest neighbors have coupling strength~$\eta$. As to the nearest-neighbor terms, for a fixed~$x$ and fixed~$j$, there is the direct interaction with~$x'$ of strength~$1$ and there is a term of strength~$\eta^2$. Thus, upon summing, the nearest-neighbor interaction has total strength $(d-1)(1+\eta^2)$. 

Since the overall strength of the interaction is irrelevant, the ratio of the strength of the next-nearest neighbor to the nearest-neighbor couplings has to be a number of the form $\frac1{d-1}\frac\eta{1+\eta^2}$ which, in particular, permits any ratio whose absolute value is bounded by~$\frac1{2(d-1)}$.

\smallskip\noindent
(2) \textsl{Yukawa potentials}:
Reflection positivity for
the Yukawa potentials can be shown by applying the criterion from
Lemma~\ref{lemma2.1}:
Fix $\mu>0$ and let $J_{x,y} = \texte^{-\mu \|x-y\|_1}$.  Then for any
observable~$f\colon\BbbH_1\to\R$,
\begin{multline}
\label{eq4.8} 
\sum_{\begin{subarray}{c}
x\in\BbbH_1\\y\in\Z^d\smallsetminus\BbbH_1
\end{subarray}}
J_{x,y}f(x)f(\vartheta^{(1)}y) 
\\
= \sum_{\begin{subarray}{c}
x_2,\dots,x_d\in\Z
\\
y_2,\dots,y_d\in\Z
\end{subarray}}
K(x,y)
\biggl(\,\sum_{x_1 >0} \texte^{-\mu x_1}f(x)\biggr) 
\biggl(\,\sum_{y_1 > 0} \texte^{-\mu y_1}f(y)\biggr),
\end{multline}
where the operator kernel $K\colon\Z^{d-1}\to\Z^{d-1}$ is defined by~$K(x,y)=\exp\{-\mu\sum_{j=2}^d |x_j-y_j|\}$.
This operator is symmetric and diagonal in the Fourier basis; a direct calculation shows that~$K$ has only positive eigenvalues.
This means that the right-hand side is non-negative, proving condition~(3) of Definition~\ref{def2}. (The other conditions are readily checked as well.)

\smallskip\noindent
(3) \textsl{Power-laws}: 
We begin by noting that all conditions on~$J_{x,y}$
in Definition~\ref{def2} are linear in~$J_{x,y}$. Therefore,
any linear combination of reflection positive~$J_{x,y}$'s with
non-negative coefficients is also reflection positive.  In particular, if we
integrate a one parameter family of interactions against a
positive measure, the result must also be RP. Now if we let
\begin{equation}
J_{x,y} = \int_0^\infty \mu^{s-1} \texte^{-\mu|x-y|_1}\textd
\mu \qquad \text{for $s>0$},
\end{equation}
then $J_{x,y}= C(s)|x-y|_1^{-s}$ and so the power laws
are RP as well.

\smallskip
We observe that in the classics, particularly, Refs.~\cite{FILS1,FILS2}, the above types of interactions are treated and the RP properties established with all distances expressed in $\ell_2$-norms. The derivations therein all rely, to some extent, on latticization of the field-theoretic counterparts to reflection positivity which were, perhaps, better known in their heyday. Our $\ell_1$ derivations, while being a more pedestrian method of extension from~$d=1$, have the advantage that they are self-contained.
\end{remark}

\subsection{Infrared bounds}
\label{sec4.3}\noindent
Our principal reason for introducing reflection positivity is to establish an
upper bound on the two point correlation term in Theorem~\ref{Th4.1}.  
This will be achieved by invoking the connection between reflection
positivity and infrared bounds. For spin systems this connection goes back 
to Ref.~\cite{FSS} where infrared bounds were used to provide proofs of phase 
coexistence in certain  continuous-spin models at low temperature.
Here we will follow the strategy of Ref.~\cite{BC}, 
and so we will keep our discussion brief. 

In order to apply infrared bounds to the problem at hand we must
first restrict consideration to those Gibbs states with
the following two properties:

\begin{property}[Torus state]
\label{P1}  
An infinite volume Gibbs
measure~$\nu_{\beta, \bh}$ is called a \emph{torus state} if it can
be obtained as a weak limit of finite-volume states with periodic
boundary conditions. (The torus states need not correspond exactly to the values~$\beta$ and~$\bh$.)
\end{property}

\begin{property}[Block averages]
\label{P2}  
An infinite
volume Gibbs measure~$\nu_{\beta, \bh}$ is said to have \emph{block
average magnetization}~$\bm_\star$ if
\begin{equation}
\label{e4.25}
\lim_{\Lambda \uparrow \Z^d} \frac1{|\Lambda|}\sum_{x\in\Lambda}\bS_x
= \bm_\star, \qquad \nu_{\beta,\bh}\text{-almost surely.}
\end{equation}
Similarly, the measure is said to have \emph{block average energy density}~$e_\star$ if
\begin{equation}
\label{e4.26}
\lim_{\Lambda \uparrow \Z^d} \frac1{|\Lambda|}\sum_{\begin{subarray}{c}
\langle x,y\rangle\\x,y\in\Lambda
\end{subarray}}
J_{x,y}\,(\bS_x,\bS_y)
= e_\star, \qquad \nu_{\beta,\bh}\text{-almost surely.}
\end{equation}
Here in Eqs.~\twoeqref{e4.25}{e4.26} the limits are along increasing sequences of square boxes centered at the origin.
\end{property}

It is conceivable that not every (extremal) Gibbs state will obey these restrictions, so the
reader might wonder how we are going to detect the desired phase transitions.
We will use an approximation argument which goes back to Ref.~\cite{BC}.
Recall the definition of the set~$\scrM_\star(\beta,\bh)$ of ``extremal magnetizations'' 
from the paragraph before Theorem~\ref{thm2.2}. Then we have:

\begin{lemma}
\label{L4.2}
For all $\beta>0$, $\bh\in\R^n$ and all $\bm_\star\in\scrM_\star(\beta,\bh)$, there exists an infinite volume Gibbs state~$\nu_{\beta,\bh}$ for interaction \eqref{1.1} which obeys Properties~\ref{P1} and~\ref{P2}.
\end{lemma}

\begin{proofsect}{Proof}
This is, more or less, Corollary~3.4 from Ref.~\cite{BC} enhanced to include the block average energy density.
\end{proofsect}

Our next goal is to show that the right-hand side of Eq.~\eqref{eq4.1} can be controlled for any Gibbs state satisfying  Properties~\ref{P1} and~\ref{P2}. To this end let $D^{-1}(x,y)$ denote the inverse of the (weighted)  Dirichlet lattice
Laplacian defined using the~$J_{x,y}$'s. Explicitly, we have
\begin{equation}
\label{eq4.11} 
D^{-1}(x,y) = \int_{[-\pi,\pi]^d} \frac{\textd k}{(2\pi)^d} \frac {\texte^{\texti k\cdot(x-y)}}{1-\hat{J}(k)},
\end{equation}
where $\hat{J}(k)=\sum_{x\in\Z^d}J_{0,x}\texte^{\texti k\cdot x}$. 
We will always work under the conditions for which the integral is convergent. Our principal estimate is now as follows:

\begin{lemma}[Infrared bound]
\label{L4.1} 
Assume that $k\mapsto (1-\hat{J}(k))^{-1}$ is Riemann
integrable.  Fix $\beta>0$, $\bh\in\R^n$ and let~$\nu_{\beta,\bh}$
be an infinite-volume Gibbs measure for interaction \eqref{1.1} that satisfies Properties~\ref{P1} and~\ref{P2}.
Let $\langle - \rangle_{\beta,\bh}$ denote the expectation with
respect to~$\nu_{\beta,\bh}$ and let~$n$ be the dimension of the underlying spin space. Then the bound
\begin{equation}
\label{eq4.14} 
\sum_{x,y \in \Z^d} v_x\bar v_y 
\bigl\langle(\bS_x - \bm_\star, \bS_y-\bm_\star)\bigr\rangle_{\beta,\bh} 
\leq \frac n\beta \sum_{x,y\in\Z^d} v_x \bar v_y D^{-1}(x,y)
\end{equation}
holds for all $v\colon\Z^d\mapsto\C$ such that $\sum_{x\in \Z^d} |v_x| < \infty$.
\end{lemma}

\begin{proofsect}{Proof}
As this lemma and its proof are similar to Lemma 3.2 of Ref.~\cite{BC}
we will stay very brief.  Let $J^{(L)}_{x,y}$ denote the
periodized interactions corresponding to the torus $\T_L$ and let
\begin{equation}
\T_L^\star =\Bigl\{\Bigl(\frac{2\pi}{L}n_1, \dots, \frac{2\pi}{L}n_d\Bigr)\colon 1 \leq n_i \leq L \Bigr\}
\end{equation} 
be the reciprocal torus. It is easy to see that the $k$-th Fourier component $\hat J^{(L)}(k)$ of the $J^{(L)}_{x,y}$'s satisfies $\hat J^{(L)}(k)=\hat J(k)$ for all~$k\in\T_L^\star$. This means that the inverse Dirichlet Laplacian on~$\T_L$ can be written in terms of the original coupling constants, i.e.,
\begin{equation}
D_L^{-1}(x,y) = \frac1{|\T_L^\star|} 
\sum_{k\in\T_L^\star\smallsetminus\{0\}} 
\frac{\texte^{\texti k\cdot(x-y)}}{1-\hat J(k)}.
\end{equation}
The infrared bound of Ref.~\cite{FILS1} then says that, for any Gibbs state~$\langle-\rangle_{\beta,\bh}^{(L)}$ on~$\T_L$ we have
\begin{equation}
\label{eq4.13a} 
\sum_{x,y \in \Z^d}\bigl\langle(\bw_x,\bS_x)(\bar\bw_y,\bS_y)\bigr\rangle_{\beta, \bh}^{(L)} 
\leq \frac 1\beta\sum_{x,y \in \Z^d} (\bw_x,\bar\bw_y) D_L^{-1}(x,y)
\end{equation}
for any absolutely summable collection of complex vectors~$(\bw_x)_{x\in\T^L}$ with~$\RE\bw_x,\IM\bw_x\in\R^n$ and~$\sum_{x\in\T^L}\bw_x=0$.

Now let us consider a torus state~$\nu_{\beta,\bh}$ with almost-surely constant block magnetization. We will first prove that~$\nu_{\beta,\bh}$ satisfies the~$L\to\infty$ version of Eq.~\eqref{eq4.13a}. By the assumption on the Riemann integrability of $\frac1{1-\hat{J}(k)}$,
\begin{equation}
\label{eq4.12} 
D_L^{-1}(x,y)\,\underset{L \longrightarrow
\infty}{\longrightarrow}\, D^{-1}(x,y),
\end{equation}
independently of~$x,y$.  Letting all~$\bw_x$ be parallel, i.e.,~$\bw_x=w_x\hate$, where~$\hate$ is a unit vector in~$\R^n$, and passing to the limit~$L\to\infty$, we thus get
\begin{equation}
\label{eq4.13} 
\sum_{x,y \in \Z^d} w_x\bar w_y\bigl\langle(\bS_x,\bS_y)\bigr\rangle_{\beta, \bh} 
\leq \frac n\beta\sum_{x,y \in \Z^d} w_x \bar w_y D^{-1}(x,y)
\end{equation}
whenever $w\colon\Z^d\to\C$ is absolutely summable and $\sum_{x\in\Z^d}w_x=0$.

In order to make the~$\bm_\star$'s appear explicitly on the left-hand side, we need to relax the condition on the total sum of the~$w_x$'s. Under the condition in Property~\ref{P2}, this is done exactly as in Lemma~3.2 of Ref.~\cite{BC}.
\end{proofsect}

\subsection{Actual proofs}
\label{sec4.4}\noindent
A key consequence of the infrared bound is the following estimate on the variance of the quantity
$\bm_0 = \sum_{x\in \Z^d} J_{0,x}\bS_x$:

\begin{lemma}[Variance bound]
\label{lemma-key}
Consider a collection $(J_{x,y})$ of coupling constants that are RP and obey Eq.~\eqref{Norm}, and let~$\scrI$ be the integral in Eq.~\eqref{2.5}. Let~$\langle-\rangle_{\beta,\bh}$ be a translation and rotation invariant Gibbs state satisfying Properties~\ref{P1} and~\ref{P2} and let~$\bm_\star=\langle\bS_0\rangle_{\beta,\bh}$. Then
\begin{equation}
\label{eq4.23}
\beta\,\bigl\langle\,|\bm_0-\bm_\star|^2\bigr\rangle_{\beta,\bh}\le n\scrI.
\end{equation}
\end{lemma}

\begin{proofsect}{Proof}
We have to show how the bound~\eqref{eq4.14} is used to estimate the variance of~$\bm_0$.  Let $(v_x)$ be defined by $v_x = J_{0,x}$. Using Lemma~\ref{L4.1} and Lemma~\ref{L4.2}, for any
$\langle-\rangle_{\beta, \bh}$ as above, this choice of the $v_x$'s leads to the variance of~$\bm_0$ on the left-hand side of Eq.~\eqref{eq4.14}, while on the right-hand side the sum turns into the integral~$\scrI$.
\end{proofsect}

The proof of Theorem~\ref{thm2.2} is now reduced to two lines:

\begin{proofsect}{Proof of Theorem~\ref{thm2.2}}
Combining Lemmas~\ref{L4.2} and~\ref{lemma-key} with Eqs.~\eqref{eq4.11b} and~\eqref{eq4.1}, we obtain Eqs.~\twoeqref{2.4}{2.5}.
\end{proofsect}

Armed with the conclusions of Theorem~\ref{thm2.2}, we can now finish also the proof of Theorem~\ref{T-philosophy}:

\begin{proofsect}{Proof of Theorem~\ref{T-philosophy}}
In light of the previous derivations, the claims in Theorem~\ref{T-philosophy} are hardly surprising. The difficulty to be overcome is the fact that the limits in Eqs.~\twoeqref{ideol1}{ideol2} are claimed for sequences of \emph{any} states, regardless of whether they obey Properties~\ref{P1} and~\ref{P2} above. 

We begin with the proof of part~(1); namely, Eq.~\eqref{ideol1}. Since~$\bm$ is the unique minimizer of~$\Phi_{\beta,\bh}$, for each~$\epsilon>0$ there exists~$\delta>0$ such that
\begin{equation}
%\label{}
\bigl\{\bm'\in\Conv(\Omega)\colon\Phi_{\beta,\bh}(\bm')<F_\MF(\beta,\bh)+\delta\bigr\}
\end{equation}
is contained in a ball $\UU_\epsilon(\bm)$ of radius~$\epsilon$ centered at~$\bm$. By~Eq.~\eqref{2.4}, once~$\beta n\frac\kappa2\scrI\le\delta$, all of~$\scrM_\star(\beta,\bh)$ must be contained in this ball. But,~$\scrM_\star(\beta,\bh)$ is the set of extremal magnetizations, and any magnetization~$\bm'$ that can be achieved in a translation-invariant state is thus in the convex hull of~$\scrM_\star(\beta,\bh)$. It follows that $\bm'\in\UU_\epsilon(\bm)$, proving Eq.~\eqref{ideol1}.

To prove Eq.~\eqref{ideol2}, let~$[e_\star,\bm_\star]$ be an extremal pair in~$\scrK_\star(\beta,\bh)$. (See the discussion prior to Theorem~\ref{thm2.2} for the definition of these objects.) Let~$\langle-\rangle_{\beta,\bh}$ be a translation and rotation invariant state for which
\begin{equation}
%\label{}
e_\star=\bigl\langle(\bS_0,\tfrac\beta2\bm_0+\bh)\bigr\rangle_{\beta,\bh}
\quad\text{and}\quad
\bm_\star=\langle\bS_0\rangle_{\beta,\bh}
\end{equation}
and suppose the state satisfies Properties~\ref{P1} and~\ref{P2}. (The existence of such a state is guaranteed by Lemma~\ref{L4.2}.) Combining Eqs.~\eqref{eq4.11b} and~\eqref{eq4.23}, we get
\begin{equation}
\label{eq4.11c}
0\le \bigl\langle(\bS_0,\bm_0)\bigr\rangle_{\beta,\bh}-|\bm_\star|^2\le\kappa n\scrI,
\end{equation}
and so, invoking the result of part~(1) of this theorem,~$e_\star$ is close to~$E(\bm_\star)$ once~$\scrI$ is sufficiently small. But this is true for all extremal pairs in~$\scrK_\star(\beta,\bh)$ and so it must be true for \emph{all} pairs in~$\scrK_\star(\beta,\bh)$. Hence,~$\scrK_\star(\beta,\bh)$ shrinks to a single point as~$\scrI\downarrow0$, which is what is claimed in part~(2) of the theorem.

To conclude the proof of the theorem, we need to show that the spin configuration converges in distribution to a product measure. Applying the DLR conditions, the conditional distribution of~$\bS_0$ given a spin configuration in~$\Z^d\setminus\{0\}$ is
\begin{equation}
\label{}
\texte^{(\bS_0,\beta\bm_0+\bh)-G(\beta\bm_0+\bh)}\mu_0(\textd\bS_0),
\end{equation}
i.e., the distribution of~$\bS_0$ depends on the rest of the spin configuration only via~$\bm_0=\sum_{x\in\Z^d}J_{0,x}\bS_x$. Hence, it clearly suffices to show that~$\bm_0$ converges to~$\bm$---the unique minimizer of~$\Phi_{\beta,\bh}$---in probability. But this is a direct consequence of the convexity bound on the left-hand side of Eq.~\eqref{eq4.11b} which tells us that, once the magnetization and energy density converge to their mean-field values, the variance of~$\bm_0$ tends to zero.
\end{proofsect}

While we cannot generally prove that, in systems with interaction~\eqref{1.1} the magnetization increases with~$\beta$, the estimates in the previous proof provide a bound on how bad the non-monotonicity can be:

\begin{lemma}[Near monotonicity of magnetization]
\label{lemma4.8} 
Let $(J_{x,y})$ be coupling constants that are RP and obey Eq.~\eqref{Norm}, and let~$\scrI$ be the integral in Eq.~\eqref{2.5}. Let~$\beta<\beta'$ and let~$\bm_\star\in\scrM_\star(\beta,\bh)$ and~$\bm_\star'\in\scrM_\star(\beta',\bh)$. Then we have:
\begin{equation}
%\label{}
|\bm_\star|^2\le|\bm_\star'|^2+\kappa n\scrI.
\end{equation}
\end{lemma}

\begin{proofsect}{Proof}
Let~$\langle-\rangle_{\beta,\bh}$ and~$\langle-\rangle_{\beta',\bh}$ be (translation and rotation invariant) states satisfying Properties~\ref{P1} and~\ref{P2} in which the above magnetizations are achieved. (Such states exist by Lemma~\ref{L4.2}.) By Eq.~\eqref{eq4.11b} we have
\begin{equation}
%\label{}
\bigl\langle(\bS_0,\bm_0)\bigr\rangle_{\beta,\bh}\ge|\bm_\star|^2,
\end{equation}
and Eqs.~\eqref{eq4.11b} and~\eqref{eq4.11c} yield
\begin{equation}
%\label{}
\bigl\langle(\bS_0,\bm_0)\bigr\rangle_{\beta',\bh}\le|\bm_\star'|^2+\kappa n\scrI.
\end{equation}
But the quantities on the left are, more or less, derivatives of the physical free energy with respect to~$\beta$ (in the parametrization introduce in Eq.~\eqref{1.1}). Hence, standard convexity arguments give us
\begin{equation}
%\label{}
\bigl\langle(\bS_0,\bm_0)\bigr\rangle_{\beta',\bh}
\ge\bigl\langle(\bS_0,\bm_0)\bigr\rangle_{\beta,\bh}.
\end{equation}
Combining these inequalities the claim follows.
\end{proofsect}

\subsection{Bounds for specific interactions}
\label{sec4.5}\noindent
Having presented the main theorem, we now argue that by
appropriately adjusting the parameters~$\mu$ and~$s$ in the Yukawa
and power law terms of an interaction, one can make the integral~$\scrI$ 
as small as desired. We begin with a general criterion along these lines:

\begin{proposition}
\label{prop4.7}
Let~$(J_{x,y}^{(\lambda)})$ be a family of translation and reflection-invariant couplings depending on a parameter~$\lambda$. Assume that the~$J_{x,y}^{(\lambda)}$ obey Eq.~\eqref{Norm} and let~$\hat J_\lambda(k)=\sum_{x\in\Z^d}J_{0,x}^{(\lambda)}\texte^{\texti k\cdot x}$ be the Fourier components. Suppose that the following two conditions are true:
\settowidth{\leftmargini}{(111i)}
\begin{enumerate}
\item[(1)]
There exists a~$\delta>0$ and a constant~$C>0$ such that for all sufficiently small~$\lambda$, we have
\begin{equation}
\label{4.40}
\frac{1-\hat J_\lambda(k)}{|k|^{d-\delta}}\ge C,\qquad k\in[-\pi,\pi]^d\setminus\{0\}.
\end{equation}
\item[(2)]
The~$\ell^2$-norm of~$(J_{0,x}^{(\lambda)})$ tends to zero as~$\lambda\to0$, i.e.,
\begin{equation}
\label{4.41}
\lim_{\lambda\to0}\sum_{x\in\Z^d}\bigl[J_{0,x}^{(\lambda)}\bigr]^2=0.
\end{equation}
\end{enumerate}
Then we have:
\begin{equation}
%\label{}
\lim_{\lambda\to0}\,\int_{[-\pi,\pi]^d}\frac{\textd k}{(2\pi)^d}
\frac{|\hat J_\lambda(k)|^2}{1-\hat J_\lambda(k)}=0.
\end{equation}
\end{proposition}

\begin{proofsect}{Proof}
Note that, by Eq.~\eqref{Norm} and condition~(1) above we have~$\hat J_\lambda(0)=1$ and~$\hat J_\lambda(k)<1$ for all~$k\ne0$. (The reflection invariance guarantees that~$\hat J_\lambda$ is an even and real function of~$k$.) First we will bound the part of the integral corresponding to $k\approx0$. To that end we pick~$r>0$ and estimate
\begin{equation}
\label{4.32a}
\int_{|k|<r}\frac{\textd k}{(2\pi)^d}
\frac{|\hat J_\lambda(k)|^2}{1-\hat J_\lambda(k)}
\le
\int_{|k|<r}\frac{\textd k}{(2\pi)^d}
\frac{1}{C|k|^{d-\delta}}=C_1r^\delta,
\end{equation}
where~$C_1=C_1(\delta,d,C)<\infty$.
Next we will attend to the rest of the integral. Let~$M(r)$ be the supremum of~$(1-\hat J_\lambda(k))^{-1}$ over all~$k\in[-\pi,\pi]^d$ with~$|k|\ge r$. By condition~(1) above, we have that $M(r)\le\frac1C r^{\delta-d}$. Therefore,
\begin{equation}
%\label{}
\int_{\begin{subarray}{c}
k\in[-\pi,\pi]^d\\|k|\ge r
\end{subarray}}
\frac{\textd k}{(2\pi)^d}
\frac{|\hat J_\lambda(k)|^2}{1-\hat J_\lambda(k)}
\le M(r)\sum_{x\in\Z^d}\bigl[J_{0,x}^{(\lambda)}\bigr]^2,
\end{equation}
where we also used Parseval's identity. By condition~(2) above, this vanishes as $\lambda\to0$, while the integral in \eqref{4.32a} can be made as small as desired by letting~$r\downarrow0$. From here the claim follows.
\end{proofsect}

Now we apply the above lemma to our specific interactions. We begin with the Yukawa potentials:

\begin{lemma}
%\label{lemma}
Let~$(J_{x,y}^{(\mu)})$ be the Yukawa interactions with parameter~$\mu$---as described in Sect.~\ref{sec1.2}---and suppose these are adjusted so that Eq.~\eqref{Norm} holds. Then $(J_{x,y}^{(\mu)})$ obey conditions~(1) and~(2) of Proposition~\ref{prop4.7} as~$\mu\downarrow0$ with~$\delta=d-2$. Consequently, in dimensions~$d\ge3$, the corresponding integral in Eq.~\eqref{2.5} tends to zero as $\mu\downarrow0$.
\end{lemma}

\begin{proofsect}{Proof}
Let $(J_{x,y}^{(\mu)})$ be as above and let~$\hat J_\mu$ denote the Fourier transform. In order to handle the overall normalization effectively, we introduce the quantity $C_\mu$ by $C_\mu\mu^d\sum_{x\ne0}\texte^{-\mu|x|_1}=1$ and note that~$C_\mu$ converges to a finite and positive limit as~$\mu\downarrow0$. From here we check that the~$\ell^2$-norm in Eq.~\eqref{4.41} scales as~$\mu^d$ and so condition~(2) of Proposition~\ref{prop4.7} follows.

It remains to prove that~$1-\hat J_\mu(k)$ is bounded from below by a positive constant times~$|k|^2$, where~$|k|$ denotes the~$\ell^2$-norm of~$k$. First we claim that for all~$\eta>0$ there exists a constant~$A<\infty$ such that for all~$k\in[-\pi,\pi]^d$,
\begin{equation}
\label{4.45}
\hat J_\mu(k)\le 1-\eta,\qquad |k|\ge A\mu.
\end{equation}
Indeed, an explicit calculation gives us
\begin{equation}
\label{4.46}
\hat J_\mu(k)=\mu^d C_\mu\sum_{x\ne0}\texte^{-\mu|x|_1+\texti k\cdot x}
\le\mu^dC_\mu\prod_{j=1}^d\Bigl\{\RE\frac1{1-\texte^{-\mu+\texti k_j}}\Bigr\},
\end{equation}
where we first neglected the condition~$x\ne0$, then wrote the result as the product over lattice directions and, finally, threw away some negative constants from each term in the product (the real parts are positive). Introducing the abbreviations $a=\texte^{-\mu}$, $\epsilon=1-a$ and~$\Delta_j=1-\cos(k_j)$, the $\epsilon$-multiple of the~$j$-th term in the product is now
\begin{equation}
%\label{}
\epsilon\,\RE\frac1{1-\texte^{-\mu+\texti k_j}}
=\frac{\epsilon^2+a\Delta_j\epsilon}{\epsilon^2+2a\Delta_j}.
\end{equation}
Now if~$\epsilon^2\ge\Delta_j$ the right-hand side is less than $1+a\epsilon$, while if~$\epsilon^2\le\Delta_j$, then it is less than~$\epsilon+\frac1{2a}\frac{\epsilon^2}{\Delta_j}$, which is $\ll1$ once~$\epsilon^2\ll\Delta_j$. Going back to Eq.~\eqref{4.46}, if at least one component of~$k$ exceeds large constant times~$\mu$ (which is itself of order~$\epsilon$), then the right-hand side of Eq.~\eqref{4.46} is small. This proves Eq.~\eqref{4.45} for~$\mu$ small; for all other~$\mu$ this holds existentially.

The condition \eqref{4.45} implies Eq.~\eqref{4.41} for~$|k|\ge A\mu$. As for the complementary values of~$k$, here we pick a small number~$\theta$ and write
\begin{equation}
\label{4.48}
1-\hat J_\mu(k)\ge C_\mu\mu^d\sum_{\begin{subarray}{c}
x\ne0\\|x|_1\le\theta/\mu
\end{subarray}}
\texte^{-\mu|x|_1}[1-\cos(k\cdot x)].
\end{equation}
By the fact that~$|k|\le A\mu$, the condition $|x|_1\le\theta/\mu$ (with~$\theta$ sufficiently small) implies that~$1-\cos(k\cdot x)\ge c(k\cdot x)^2$ for some~$c>0$. Plugging this into Eq.~\eqref{4.48} and using that the domain of the sum is invariant under reflection of any component of~$x$, the result will be proportional to~$|k|^2$. The constant of proportionality is of order~$\mu^{-2}$ and so condition~(1) is finally proved.
\end{proofsect}

Next we attend to the power laws:

\begin{lemma}
%\label{lemma}
Let~$(J_{x,y}^{(s)})$ be the power-law interactions with exponent~$s>d$---see Sect.~\ref{sec1.2}---and suppose these are adjusted so that Eq.~\eqref{Norm} holds. Then $(J_{x,y}^{(s)})$ obey conditions~(1) and~(2) of Proposition~\ref{prop4.7} as~$s\downarrow d$ with any~$\delta<d$. Consequently, the corresponding integral in Eq.~\eqref{2.5} tends to zero as $s\downarrow d$ in all~$d\ge1$.
\end{lemma}

\begin{proofsect}{Proof}
Our first item of business will again be the overall normalization. Let~$C_s$ be the constant defined by
\begin{equation}
%\label{}
C_s(s-d)\sum_{x\ne0}|x|_1^{-s}=1. 
\end{equation}
As is not hard to check,~$C_s$ tends to a positive and finite limit as~$s\downarrow d$. Since~$\sum_{x\ne0}|x|_1^{-2s}$ is uniformly bounded for all~$s>d$, the~$\ell^2$-norm in Eq.~\eqref{4.41} is proportional to~$(s-d)$. This proves condition~(2) of Proposition~\ref{prop4.7}. 

In order to prove condition~(1), we first write
\begin{equation}
%\label{}
1-\hat J_s(k)=C_s(s-d)\sum_{x\ne0}|x|_1^{-s}\bigl(1-\cos(k\cdot x)\bigr),
\end{equation}
where~$\hat J_s$ is the Fourier transform of the~$(J_{x,y}^{(s)})$. Consider the set~$\RR_k=\{x\in\Z^d\colon \cos(k\cdot x)\le0\}$, which we note is the union of strips of width---and separation---of the order~$O(1/|k|)$ which are perpendicular to vector~$k$. A simple bound gives us
\begin{equation}
\label{4.54}
\sum_{x\ne0}|x|_1^{-s}\bigl(1-\cos(k\cdot x)\bigr)\ge\sum_{x\in\RR_k}|x|_1^{-s}.
\end{equation}
Next we let~$\RR_k'=\{x\in\Z^d\colon |x\cdot k|>\pi\}$. The fact that~$|x|_1^{-s}$ decreases with distance allows us to bound the second sum in Eq.~\eqref{4.54} by a similar sum with~$x\in\RR_k'$. Using the usual ways to bound sums by integrals, we thus get
\begin{equation}
%\label{}
1-\hat J_s(k)\ge C(s-d)\int_{|k\cdot x|\ge\pi}\frac{\textd x}{|x|^s},
\end{equation}
where~$C$ is a positive constant (independent of~$s$) and~$|x|$ is the~$\ell^2$-norm of~$x$. Extracting a factor of~$|k|^{s-d}$, the resulting integral \emph{times} $(s-d)$ is uniformly positive for all~$s>d$. Hence we proved that for some~$c'>0$,
\begin{equation}
%\label{}
1-\hat J_s(k)\ge c'|k|^{s-d}
\end{equation}
for all~$s>d$ and all~$k\in[-\pi,\pi]^d$, and so condition~(1) of Proposition~\ref{prop4.7} holds as stated.
\end{proofsect}

\section{Proofs: Mean-field theories}
\label{sec5}

\subsection{Blume-Capel model}
\label{sec5.1}\noindent
We begin by giving the proof of Theorem~\ref{BCT1} which deals with the mean-field theory of the Blume-Capel model. The core of this proof, and other proofs in this paper, are certain facts about the mean-field theory of the Ising model in an external field. In the formalism of Sect.~\ref{sec2.2}, this model corresponds to the~$q=2$ Potts model.  The magnetizations are parameterized by a pair of quantities~$(z_1,z_{-1})$, where~$z_1+z_{-1}=1$, which represent the mole-fractions of plus and minus spins. The mean-field free energy is given by
\begin{equation}
\label{IsingFE}
\Phi_{J,h}=Jz_1z_{-1}-hz_1+z_1\log z_1+z_{-1}\log z_{-1}.
\end{equation}
The following properties are the results of straightforward calculations:
\settowidth{\leftmargini}{(1111)}
\begin{enumerate}
\item[(I1)]
If~$h=0$ and~$J\le2$, then the only local---and global---minimum occurs at~$z_1=z_{-1}$.
\item[(I2)]
If~$h=0$ and~$J>2$, then there is only one local minimum with~$z_1\ge z_{-1}$ and it satisfies~$Jz_1>1>Jz_{-1}$. A corresponding local minimum with with~$z_1\ge z_{-1}$ exists and obeys~$Jz_{-1}>1>Jz_1$.
\item[(I3)]
Let now~$h$ be arbitrary. If~$(z_1,z_{-1})$ is a local minimimum of~$\Phi_{J,h}$, then~$m=z_1-z_{-1}$ satisfies~$J(1-m^2)\le1$.
\end{enumerate}
These properties are standard; for some justification see, e.g., the proof of Lemma~4.4 in Ref.~\cite{BC}.

\begin{proofsect}{Proof of Theorem~\ref{BCT1}}
Let~$(x_1,x_0,x_{-1})$ be a triplet of positive variables which corresponds to a local minimum of the Blume-Capel free-energy function $\Phi_{\beta,\lambda}$ from Eq.~\eqref{BC1.1}. A simple calculations shows that the derivative of the entropy part of $\Phi_{\beta,\lambda}$ is singular in the limit when any component of~$(x_1,x_0,x_{-1})$ tends to zero, while nothing spectacular happens to the energy. Therefore, the minimum must lie strictly inside the simplex of allowed values. Accounting for the constraint $x_1 + x_0 + x_{-1} = 1$, the condition that the gradient of~$\Phi_{\beta,\lambda}$ vanish at~$(x_1,x_0,x_{-1})$ translates into the equations \eqref{3.15a}.

Due to the symmetry between~$x_1$ and~$x_{-1}$, we may (and will) assume for simplicity that~$x_1\ge x_{-1}$. First we claim that, under this condition, we have~$4\beta x_{-1}\le1$. Indeed, for a fixed~$x_0$, the Blume-Capel mean-field free energy~$\Phi_{\beta,\lambda}$ expressed in terms of~$(z_1,z_{-1})$, where~$z_{\pm1}=x_{\pm1}/(1-x_0)$, is proportional to the Ising free energy~\eqref{IsingFE} with $J=4\beta(1-x_0)$. Since the Ising pair~$(z_1,z_{-1})$ is at its local minimum, we have $Jz_{-1}=4\beta x_{-1}\le1$ by property~(I2) above.

Once we know that~$x_{-1}$ is small, the question is whether~$x_0$ and~$x_1$ divide the amount $1-x_{-1}$ democratically or autocratically. Here we observe that, once again, for a fixed~$x_{-1}$, the~$(x_1,x_0)$-portion of the Blume-Capel mean-field free energy~$\Phi_{\beta,\lambda}$ is proportional to its Ising counterpart in Eq.~\eqref{IsingFE} with $J=\beta(1-x_{-1})$ and~$h=3\beta x_{-1}-\lambda$. In light of property~(I3) above, the magnetization variable~$m=(x_1-x_0)/(1-x_{-1})$ thus satisfies the bound $J(1-m^2)\le1$. Using the inequality~$\sqrt{1-a}\ge1-a$ valid for all~$a\le1$, we have
\begin{equation}
%\label{}
\frac{|x_1-x_0|}{1-x_{-1}}\ge1-\frac1{\beta(1-x_{-1})}
\end{equation}
once~$\beta$ is sufficiently large. Some simple algebra now shows that this implies
\begin{equation}
%\label{}
2\beta\min\{x_1,x_0\}\le1.
\end{equation}
Using these findings in Eq.~\eqref{3.15a} and extracting appropriate inequalities we derive the bounds listed in~(1) and~(2) with~$C$ being a numerical constant.

To derive the asymptotics \eqref{3.16b} on the free-energy gap for~$\lambda\approx0$, let us first evaluate the free energy at a generic local minimum. Suppose~$(x_1,x_0,x_{-1})$ obey Eq.~\eqref{3.15a} and let~$\varTheta$ denote the logarithm of the quantity in Eq.~\eqref{3.15a}. A direct calculation shows that then
\begin{equation}
%\label{}
\Phi_{\beta,h}=-4\beta x_1x_{-1}+\beta x_0^2+\varTheta.
\end{equation}
Now let us consider a minimum with~$x_0$ dominant. Then the inequality~$\beta(1-x_0)=\beta(x_1+x_{-1})\le\ffrac34<1$ shows that the~$(x_1,x_{-1})$ Ising pair is subcritical. By~(I1) above we must have~$x_1=x_{-1}=\frac12(1-x_0)$ and, as is seen by a direct calculation,~$x_0$ can be determined from the equation
\begin{equation}
%\label{}
\frac{1-x_0}{x_0}=2\texte^{-\beta+\lambda}.
\end{equation}
In particular, for~$\lambda$ bounded we have~$1-x_0=2\texte^{-\beta+\lambda}+O(\texte^{-2\beta})$. 
Similarly, if the minimum corresponds to a triple dominated by~$x_1$, our bounds show that~$x_0=1-x_1+O(\texte^{-4\beta})$ and so we have
\begin{equation}
%\label{}
x_1=\bigl(1-x_1+O(\texte^{-4\beta})\bigr) \texte^{\beta+\lambda+O(\beta \texte^{-\beta})}.
\end{equation}
From here we have~$1-x_1=\texte^{-\beta-\lambda}+O(\beta \texte^{-2\beta})$.

Now we are ready to derive Eq.~\eqref{3.16b}. First, using that~$\varTheta=\log x_0+\beta(1-2x_0)+\lambda$ we have
\begin{multline}
\label{3.21a}
\qquad
\phi_0(\beta,\lambda)=-4\beta x_1x_{-1}+\beta (1-x_0)^2+\lambda+\log x_0
\\
=\lambda-2\texte^{-\beta+\lambda}+O(\texte^{-2\beta}).
\qquad
\end{multline}
Next, in light of~$\varTheta=\log x_1+4\beta x_{-1}$ and the bounds proved on~$x_{-1}$ in~(2) above we have
\begin{multline}
\label{3.22a}
\qquad
\phi_1(\beta,\lambda)=-4\beta x_1x_{-1}+\beta x_0^2+\log x_1+4\beta x_{-1}
\\
=-\texte^{-\beta-\lambda}+O(\beta \texte^{-2\beta}).
\qquad
\end{multline}
Combining Eqs.~\twoeqref{3.21a}{3.22a}, the desired relation \eqref{3.16b} is proved.
\end{proofsect}

We finish this section with a computational lemma that will be useful in the proof of Theorem~\ref{BCT2}:

\begin{lemma}
\label{lemma-BCmax}
There exists~$\alpha>0$ and, for each~$C\gg1$, there exists $\beta_0<\infty$ such that the following is true for all~$\beta\ge\beta_0$ and all~$\lambda$ with $|\lambda|\le C\texte^{-\beta}$: If~$(x_1,x_0,x_{-1})$ is a triplet with
\begin{equation}
\label{5.9a}
\max\{x_1,x_0,x_{-1}\}=1-C\texte^{-\beta},
\end{equation}
then
\begin{equation}
\label{eq5.10}
\Phi_{\beta,\lambda}(x_1,x_0,x_{-1})-\inf\Phi_{\beta,\lambda}\ge \alpha(C\log C)\texte^{-\beta}.
\end{equation}
Here~$\Phi_{\beta,\lambda}$ is the function in Eq.~\eqref{BC1.1} and $\inf\Phi_{\beta,\lambda}$ is its absolute minimum.
\end{lemma}

\begin{proofsect}{Proof}
An inspection of Eqs.~\twoeqref{3.21a}{3.22a} shows that, once~$|\lambda|\le C\texte^{-\beta}$, we have that $|\inf\Phi_{\beta,\lambda}|$ is proportional to~$C\texte^{-\beta}$ and so we just have to prove that, once~$C$ is sufficiently large, $\Phi_{\beta,\lambda}(x_1,x_0,x_{-1})$ is proportional to~$(C\log C)\texte^{-\beta}$. We will focus on the situation when the maximum in Eq.~\eqref{eq5.10} is achieved by~$x_1$; the other cases are handled similarly. 

By our assumption we have that~$x_0$ and~$x_{-1}$ are quantities less than~$C\texte^{-\beta}$. Inspecting the various terms in Eq.~\eqref{BC1.1}, we thus have
\begin{equation}
\begin{aligned}
\beta x_0(1-x_0)&=\beta x_0+O(\beta C^2\texte^{-2\beta}),
\\
\beta x_1x_{-1}&=4\beta x_{-1}+O(\beta C^2\texte^{-2\beta}),
\\
x_1\log x_1&=-C\texte^{-\beta}+O(C^2\texte^{-2\beta}),
\end{aligned}
\end{equation}
Plugging these back into the definition of~$\Phi_{\beta,\lambda}$ we get
\begin{multline}
\label{eq5.11}
\qquad
\Phi_{\beta,\lambda}(x_1,x_0,x_{-1})=x_0[\beta+\log x_0]+x_{-1}[4\beta+\log x_{-1}]
\\
+\lambda x_0-C\texte^{-\beta}+O(\beta C^2\texte^{-2\beta}).
\qquad
\end{multline}
Now~$|\lambda x_0|\le|\lambda|\le C\texte^{-\beta}$, and if~$\beta_0$ is such that $\beta C\texte^{-\beta}\ll1$, the last three terms on the right-hand side are all of order~$C\texte^{-\beta}$. It thus suffices to to prove that the first two terms exceed a constant times~$(C\log C)\texte^{-\beta}$.

We first replace~$4\beta$ by~$\beta$ in Eq.~\eqref{eq5.11} and then substitute~$x_0=y_0\texte^{-\beta}$ and~$x_{-1}=y_{-1}\texte^{-\beta}$. The relevant two terms on the right-hand side then equal $\texte^{-\beta}[y_0\log y_0+y_{-1}\log y_{-1}]$. Under the condition \eqref{5.9a}---which implies that at least one of the~$y$'s is larger than~$\ffrac C2$---this is a number of order~$\texte^{-\beta}C\log C$ (for~$C\gg1$). The right-hand side of Eq.~\eqref{eq5.11} is thus of order~$\texte^{-\beta}C\log C$ whenever~$\beta\ge\beta_0$, which proves the desired claim.
\end{proofsect}

\subsection{Potts model: Preliminaries}
\label{sec5.2}\noindent
Next we turn our attention to the mean-field theory of the Potts model. In the present section we will first establish some basic properties of the (local) minimizers of the Potts mean-field free energy. The proof of Theorem~\ref{TH1} dealing with positive fields is then the subject of Sect.~\ref{sec5.3}. The negative-field portion of our results (Theorem~\ref{TH2}) is somewhat more involved and we defer its discussion to Sect.~\ref{sec5.4}.

We invite the reader to recall the representation of magnetizations in terms of barycentric coordinates in Eq.~\eqref{1.14a}, the mean-field free-energy function~$\Phi_{\beta,h}$ from Eq.~\eqref{1.15a} and the transitional coupling~$\betaMF^{(q)}$ for the~$q$-state Potts model from Eq.~\eqref{eq1}. We begin with some general monotonicity properties of the minimizers:

\begin{lemma}[Monotonicity in~$h$]
\label{L2}
For any $\beta \geq 0$ we have:
\settowidth{\leftmargini}{(1111)}
\begin{enumerate}
\item[(1)]
Let $h<h'$, let $x_1$ be the first barycentric coordinate of a global minimum of $\Phi_{\beta,h}^{(q)}$ and let~$x_1'$ be the first barycentric coordinate of a global minimum of $\Phi_{\beta,h'}^{(q)}$.  Then $x_1\le x_1'$.
\item[(2)]
Let~$(x_1,\dots,x_q)$ be the probability vector corresponding to a global minimizer of~$\Phi_{\beta,h}^{(q)}$.
If $h>0$ then~$x_1>\max\{x_2,\dots,x_q\}$. Similarly, if~$h<0$ then~$x_1<\min\{x_2,\dots,x_q\}$.
\item[(3)]
If $h\mapsto \bm(\beta,h)$ is a differentiable trajectory of local extrema, then
\begin{equation}
%\label{}
\frac{\textd}{\textd h} \Phi_{\beta,h}^{(q)}(\bm(\beta,h)) =-x_1(\beta,h),
\end{equation}
where~$x_1(\beta,h)$ is the first component of~$\bm(\beta,h)$ in the decomposition into~$(\hatv_1,\dots,\hatv_q)$.
\end{enumerate}
\end{lemma}

\begin{proofsect}{Proof}
(1) Let~$\bm\in\Conv\Omega$. Then we have
\begin{equation}
\label{L1.1a}
\Phi_{\beta,h}^{(q)}(\bm)-\Phi_{\beta,h'}^{(q)}(\bm) = (h'-h)x_1,
\end{equation}
where~$x_1$ is the first component of~$\bm$.
Let $x_1$ and~$x_1'$ be as above and let~$\bm$ and~$\bm'$ be the corresponding minimizers.
Then Eq.~\eqref{L1.1a} implies
\begin{equation}
\label{L1.2a}
x_1\leq \frac{\Phi_{\beta,h}^{(q)}(\bm)-\Phi_{\beta,h'}^{(q)}(\bm')}{h'-h}
\end{equation}
Similar reasoning gives
\begin{equation}
\label{L1.3a}
x_1'\geq\frac{\Phi_{\beta,h}^{(q)}(\bm)-\Phi_{\beta,h'}^{(q)}(\bm')}{h'-h}.
\end{equation}
Combining Eqs.~\eqref{L1.2a} and~\eqref{L1.3a} gives the result.

(2) Let~$h>0$ and let~$(x_1,\dots,x_q)$ be a probability vector with~$x_1<x_2$. Interchanging~$x_1$ and~$x_2$ shows that, due to the interaction with the field, the~$q$-tuple~$(x_2,x_1,\dots,x_q)$ has strictly lower free energy than $(x_1,\dots,x_q)$, i.e., $(x_1,\dots,x_q)$ could not have been a global minimizer. Hence~$x_1\ge x_2$. To rule out~$x_1=x_2$ we note that~$x_1,x_2>0$ and so the gradient of the free energy, subject to the constraint $x_1+x_2=\text{const}$, must vanish. Hence~$x_1\texte^{-\beta x_1-h}=x_2\texte^{-\beta x_2}$ which forces~$x_1\ne x_2$. The cases~$h<0$ are handled similarly.

(3) This is a consequence of the fact that the gradient~$\nabla \Phi_{\beta,h}^{(q)}$ vanishes at any local extremum in the interior of~$\Conv(\Omega)$.
\end{proofsect}

\begin{lemma}[Monotonicity in~$\beta$]
\label{L1}
Fix~$h\in\R$. If $\beta\mapsto \bm(\beta,h)$ is a differentiable trajectory of local extrema, then
\begin{equation}
%\label{}
\frac{\textd}{\textd \beta} \Phi_{\beta,h}^{(q)}(\bm(\beta,h)) = -\frac12 |\bm(\beta,h)|^2.
\end{equation}
\end{lemma}

\begin{proofsect}{Proof}
The proof is analogous to that of Lemma~\ref{L2}(3).
\end{proofsect}

The next lemma significantly narrows the list of possible candidates for global minimizers:

\begin{lemma}[Symmetries of global minimizers]
\label{L3}
Let $\Phi_{\beta,h}^{(q)}(\bm)$ be the mean-field free-energy function.  Let $\bm\in\Conv\Omega$ be a global minimum of $\Phi_{\beta,h}^{(q)}$ and let~$(x_1,\dots,x_q)$ be the corresponding probability vector of barycentric coordinates.
\settowidth{\leftmargini}{(1111)}
\begin{enumerate}
\item[(1)]
If $h>0$, then
\begin{equation}
\label{1.10}
x_1> x_2=\dots=x_q.
\end{equation}
\item[(2)]
If $h<0$, then~$(x_1,\dots,x_q)$ is a permutation in indices~$x_2,\dots,x_q$ of a vector with
\begin{equation}
\label{1.11}
x_1< x_2=\dots=x_{q-1}\le x_q.
\end{equation}
\end{enumerate}
\end{lemma}

\begin{proofsect}{Proof}
The main idea of the proof is that the variables~$x_2,\dots,x_q$, properly scaled, behave like a~$(q-1)$-state, zero-field Potts model. Abusing the notation slighly, let us write $\Phi_{\beta,h}^{(q)}(x_1,\dots,x_q)$ instead of~$\Phi_{\beta,h}^{(q)}(\bm)$ whenever~$\bm$ corresponds to the probability vector $(x_1,\dots,x_q)$. In looking for global minima, we may assume that all~$x_k$'s satisfy $x_k\in(0,1)$. Letting
\begin{equation}
%\label{}
z_k=\frac{x_k}{1-x_1},\qquad k=2,\dots,q,
\end{equation}
this allows us to write
\begin{equation}
\label{E3.1}
\Phi_{\beta,h}^{(q)}(x_1,\dots,x_q) = (1-x_1)\Phi_{\beta(1-x_1),0}^{(q-1)}(z_2,\dots,z_q) + R(x_1),
\end{equation}
where $R(x_1)$ is a function of~$x_1$ (and~$\beta$ and~$h$). The rest of the proof is based on some basic properties of the zero-field Potts free energy for which we refer the reader back to Sect.~\ref{sec2.2}.

Let~$(x_1,\dots,x_q)$ correspond to a global minimum. A principal conclusion coming from Eq.~\eqref{E3.1} is that the components of the vector~$(x_2,\dots,x_q)$, ordered increasingly, satisfy $x_2=\dots=x_{q-1}\le x_q$. Using part~(2) of Lemma~\ref{L2}, this immediately implies Eq.~\eqref{1.11}. To prove Eq.~\eqref{1.10}, let $h>0$ and let $(\tilde x_1,\dots,\tilde x_q)$ be a global minimizer at zero field with maximal value of~$\tilde x_1$. By general facts about the zero-field problem, this forces~$\beta(1-\tilde x_1)<\betaMF^{(q-1)}$ and, since part~(2) of Lemma~\ref{L2} implies that~$x_1>\tilde x_1$, also $\beta(1-x_1)<\betaMF^{(q-1)}$. Hence, the variables~$(z_2,\dots,z_q)$ correspond to a subcritical Potts model and thus~$z_2=\dots=z_q$. Invoking again Lemma~\ref{L2}(2), we have Eq.~\eqref{1.10}.
\end{proofsect}

\subsection{Potts model: Positive fields}
\label{sec5.3}\noindent
Next we will focus on the cases with~$h>0$. Our first step is to characterize the local and global minima of~$\bm\mapsto\Phi_{\beta,h}^{(q)}(\bm)$ for~$\bm$ restricted to satisfy Eq.~\eqref{1.10}. While we could appeal to the ``on-axis'' formalism from Ref.~\cite{BC}, we will keep the requisite calculations more or less self-contained.

For any probability vector satisfying Eq.~\eqref{1.10}, let us consider the parametrization $\theta=\frac q{q-1}m$, where~$m$ denotes the scalar magnetization defined via~$x_1=\frac1q+m$ and~$x_k=\frac1q-\frac m{q-1}$, $k=2,\dots,q$. (The physical values of~$\theta$ are $\theta\in[0,1]$.) Let~$\phi_{\beta,h}(\theta)$ denote the value of~$\Phi_{\beta,h}^{(q)}(\bm)$ where~$\bm$ corresponds to the above~$(x_1,\dots,x_q)$. Then we have:

\begin{lemma}[``On-axis'' minima]
\label{lemma1.4}
The local minima of~$\theta\mapsto\phi_{\beta,h}(\theta)$ are solutions to the equation $\theta=f(\theta)$, where
\begin{equation}
\label{ET1.2}
f(\theta)=\frac{\texte^{\beta\theta+h} -1}{\texte^{\beta\theta+h} +q-1}.
\end{equation}
Moreover, let~$\beta_0=4\frac{q-1}q$. Then
\begin{enumerate}
\item[(1)]
For all~$\beta\le \beta_0$ and all~$h\in\R$, the equation~$\theta=f(\theta)$ has only one solution.
\item[(2)]
For~$\beta>\beta_0$ there exists an interval~$(h_-,h_+)$ such that~$\theta=f(\theta)$ has three distinct solutions once~$h\in(h_-,h_+)$ and only one solution for~$h\not\in[h_-,h_+]$. At~$h=h_\pm$, there are two distinct solutions. Once~$h\ne h_\pm$, only the extreme solutions (the largest and the smallest) correspond to local minima of~$\theta\mapsto\phi_{\beta,h}(\theta)$.
\end{enumerate}
Finally, for each~$\beta>\beta_0$, there exists a number~$h_1=h_1(\beta)\in(h_-,h_+)$ such that the global minimizer of~$\theta\mapsto\phi_{\beta,h}(\theta)$ is unique as long as~$h\ne h_1$. On the other hand, for~$h=h_1$ there are two distinct global minimizers (the two extreme solutions of~$\theta=f(\theta)$).
\end{lemma}

\begin{remark}
Although the above holds as stated in complete generality, it is only useful (in the present context) for~$\beta<\betaMF^{(q)}$. In particular, for~$\beta\ge\betaMF^{(q)}$, while $h_1(\beta)$ continues on taking negative values, it does not correspond to any equilibrium commodity.
\end{remark}

\begin{proofsect}{Proof of Lemma~\ref{lemma1.4}}
Since the derivative of~$\theta\mapsto\phi_{\beta,h}(\theta)$ diverges as~$\theta$ tends to either zero or one, all local minima will lie in~$(0,1)$. Differentiating with respect to~$\theta$ we find that these must satisfy~$f(\theta)=\theta$ with~$f$ as given above.

In order to characterize the solutions to~$\theta=f(\theta)$, let us calculate the first two derivatives of this function:
\begin{equation}
\label{ET1.5}
f'(\theta)=\beta\frac{\texte^{\beta\theta+h}}{\texte^{\beta\theta+h} +q-1}\bigl(1-f(\theta)\bigr)
\end{equation}
and
\begin{equation}
\label{ET1.8}
f''(\theta)=\beta^2\frac{\texte^{\beta\theta+h}}{\texte^{\beta\theta+h}+q-1}\bigl(1-f(\theta)\bigr)\left(1-2\frac{\texte^{\beta\theta+h}}{\texte^{\beta\theta+h} +q-1}\right).
\end{equation}
Since we also have $f(\theta)<1$, we find that~$f$ is strictly increasing, strictly convex for~$\theta<\thetaI$ and strictly concave for~$\theta>\thetaI$, where~$\thetaI$ is the inflection point of~$f$, which is given by
\begin{equation}
\label{ET1.3}
\frac{\texte^{\beta\theta+h}}{\texte^{\beta\theta+h} +q-1} = \frac12,
\end{equation}
i.e.,~$\texte^{\beta\theta+h}=q-1$. In particular, the derivative~$f'(\theta)$ is maximal at~$\theta=\thetaI$, where it equals $f'(\thetaI)=\frac \beta4\frac q{q-1}$.

Let us suppose that~$f'(\thetaI)\le1$, which is equivalent to $\beta\le\beta_0$. Then there is only one solution to~$\theta=f(\theta)$, proving~(1) above. Let us now assume that~$f'(\thetaI)>1$. The fact that increasing~$h$ amounts to ``shifting the graph of~$f$ to the left'' implies that there exists an~$h_0$ such that~$\thetaI$ solves~$\theta=f(\theta)$ for~$h=h_0$. Similar arguments show that there exists a unique value~$h_+>h_0$ such that the diagonal line (at 45$^\circ$) is tangent to the graph of~$f$ at some~$\theta<\thetaI$, and a similar value~$h_-<h_0$ such that the diagonal line is tangent to the~$\theta\ge\thetaI$ portion of the graph of~$f$. For~$h\in[h_-,h_+]$, there are altogether three solutions, labeled $\thetaL<\thetaM<\thetaU$, where~$f'(\theta)\le1$ at~$\theta=\thetaL,\thetaU$ while~$f'(\thetaM)\ge1$ (with the inequalities strict when~$h\ne h_\pm$). 

The ``dynamics'' of these solutions as~$h$ changes is easy to glean from the above picture. 
First~$\thetaL$ is defined for all~$h\le h_+$ while~$\thetaU$ is defined for all $h\ge h_-$.
Now, as~$h$ decreases through~$h_-$, the middle~$\thetaM$ and upper~$\thetaU$ solutions merge and disappear; and similarly for~$\thetaM$ and~$\thetaL$ as~$h$ increases through~$h_+$. Only the remaining solution continues to exist in the complementary part of the~$h$-axis.  Clearly, both~$\thetaL$ and~$\thetaU$ are continuous and strictly increasing on the domain of their definition with~$\thetaL\to0$ as~$h\to-\infty$ and~$\thetaU\to1$ as~$h\to\infty$. Since~$\phi_{\beta,h}(\theta)$ has local maxima at~$\theta=0$ and~$1$, we must have that~$\thetaL$ and~$\thetaU$ are local minima and~$\thetaM$ is a local maximum of $\phi_{\beta,h}$. (These are strict except perhaps at $h\ne h_\pm$.) This finishes the proof of~(2). 

It remains to prove the existence of the transitional field-strength~$h_1$. By Lemma~\ref{L3}, every global minimizer~$\bm\mapsto\Phi_{\beta,h}^{(q)}(\bm)$ corresponds to either~$\thetaL$ or~$\thetaU$. 
Observe that, since~$\thetaU$ and~$\thetaL$ never enter the portion of the graph of~$f$ where~$f'$ exceeds one, we have~$\thetaU\ge\thetaU(h_+)>\thetaL(h_-)\ge\thetaL$ and so the difference~$\thetaU-\thetaL$ is uniformly positive.
Consequently, the values~$\Phi_{\beta,h}^{(q)}$ at the corresponding magnetizations change at a strictly different rate with~$h$ (see Lemma~\ref{L2}). In particular, there exists a unique point $h_1(\beta)\in(h_-,h_+)$, where the status of the global minimizer changes from~$\thetaL$ to~$\thetaU$. By continuity, at~$h=h_1$, both one-sided limits are minimizers of~$\Phi_{\beta,h}$.
\end{proofsect}

Now we are ready to finish the prove of Theorem~\ref{TH1}.

\begin{proofsect}{Proof of Theorem~\ref{TH1}}
Most of the claims of the theorem have already been proved. Indeed, let~$h_1$ be as in Lemma~\ref{lemma1.4} and let~$\beta\ge\betaMF^{(q)}$. By the properties of the zero-field Potts model, the maximal solution to~$\theta=f(\theta)$ is a global minimizer of~$\theta\mapsto\phi_{\beta,0}(\theta)$. It follows that~$h_1(\beta)\le0$ for~$\beta\ge\betaMF^{(q)}$. Invoking also Lemma~\ref{L3}(1), we thus conclude that for~$\beta\le \beta_0$ or~$\beta\ge\betaMF^{(q)}$ and~$h>0$, the global minimizer of~$\bm\mapsto\Phi_{\beta,h}^{(q)}(\bm)$ is unique, while for~$\beta\in(\beta_0,\betaMF^{(q)})$ this is only true when~$h\ne h_1(\beta)$. This establishes parts (2) and (3) of the theorem. It thus remains to prove the strict inequality between~$x_1$ and~$x_2=\dots=x_1$ in part~(1)---the rest follows by Lemma~\ref{L3}(1)---and the properties of~$\beta\mapsto h_1(\beta)$ in part~(4).

First, it is easy to see that~$h_1$ is continuous. Indeed,  let~$\beta'\in(\beta_0,\betaMF^{(q)}]$ and suppose that~$\beta\mapsto h_1(\beta)$ has two limit points as~$\beta\to \beta'$. By a simple compactness argument, there are two distinct minimizers of~$\phi_{\beta',h}^{(q)}$ for~$h$ at these limit points, which contradicts the uniqueness of~$h_1(\beta')$. Applying this to $\beta'=\betaMF^{(q)}$, we thus have that~$h_1(\beta)\to0$ as~$\beta\to\betaMF^{(q)}$. 

Second, we claim that $\beta\mapsto h_1(\beta)$ is actually strictly decreasing. To this end, let~$\bm_+(\beta)$ and~$\bm_-(\beta)$ denote the values of the two global minimizers of~$\bm\mapsto\Phi_{\beta,h}^{(q)}(\bm)$ at~$h=h_1(\beta)$ and let~$x_1^+(\beta)$ and~$x_1^-(\beta)$ denote the corresponding first components. From Lemmas~\ref{L2} and~\ref{L1} we can now extract
\begin{equation}
\label{ET1.7}
\frac\textd{\textd \beta}h_1(\beta)=-\frac12\frac{|\bm_+(\beta)|^2-|\bm_-(\beta)|^2}{x_1^+(\beta)-x_1^-(\beta)},
\end{equation}
which the reader will note is the Clausius-Clapeyron relation.
Since both~$x_1$ and~$|\bm|$ are increasing with the scalar magnetization, the right hand side is negative and so~$\beta\mapsto h_1(\beta)$ is strictly decreasing. 

Third, we turn our attention to the inequality~$x_1>x_2=\dots=x_q$ once~$h>0$. In light of Eq.~\eqref{1.10}, it suffices to show that, for~$h>0$, the state with equal barycentric coordinates is not a local minimum once~$h>0$. This is directly checked by differentiating Eq.~\eqref{1.15a} subject to appropriate constraints. Finally, we will compute the value of~$h$ at the end of the line~$h\mapsto\beta_+(h)$. Let~$\theta_+(h)$ and~$\theta_-(h)$ denote the two distinct (extremal) solutions of~$f(\theta)=\theta$, with~$f$ as in Eq.~\eqref{ET1.2}, for~$\beta=\beta_+(h)$. As~$h$ increases, $\beta_+$ decreases to~$\beta_0$ and~$\theta_\pm$ converge to a single value~$\theta_0$---the \emph{unique} solution of~$f(\theta)=\theta$ at~$\beta=\beta_0$. But the inflection point,~$\thetaI$, is always squeezed between~$\theta_+$ and~$\theta_-$, and so we must have~$\theta_0=\thetaI$. Now the inflection point is characterized by~$\texte^{\beta\thetaI+h}=q-1$ and the equation~$\theta=f(\theta)$ gives us that~$\beta_+(h)=\beta_0$ at~$h=\hc$.
\end{proofsect}

\subsection{Potts model: Negative fields}
\label{sec5.4}\noindent
The goal of this section is to give the proof of Theorem~\ref{TH2}. The difficulty here is that, on the basis of Eq.~\eqref{1.11}, the full-blown optimization problem is intrinsically two-dimensional. We begin with some lemmas that encapsulate the computational parts of the proof. First we will address the symmetric minima by describing the solutions to the ``on-axis'' equation:

\begin{lemma}
\label{lemma5.5}
Let~$\beta\ge0$ and~$h<0$ and let~$g\colon[0,\frac1{q-1}]\to\R$ be the
function
\begin{equation}
\label{eq5.23}
g(\theta)=\frac{\texte^{\beta\theta-h}-1}{(q-1)\texte^{\beta\theta-h} +1}.
\end{equation}
Then~$g$ is increasing, concave and satisfies~$g(0)>0$
and~$g(\theta)<1$. In particular, the equation~$g(\theta)=\theta$ has a unique
solution on~$[0,\frac1{q-1}]$.
\end{lemma}

\begin{proofsect}{Proof}
This is a result of straightforward computations which are not entirely dissimilar from those in Eqs.~\twoeqref{ET1.5}{ET1.8}.
\end{proofsect}

The two-parameter nature of solutions of the form~\eqref{1.11} will be handled by fixing the first barycentric coordinate and optimizing over the remaining ones. Here the following property of the resulting ``partial minimum'' will turn out to be very useful:

\begin{lemma}
\label{lemma5.6}
Let~$\beta > \betaMF^{(q-1)}$ and let~$\tilde a$ be the minimum of $\ffrac1q$ and the
quantity~$a$ satisfying $\beta(1-a)=\betaMF^{(q-1)}$. For
each~$x\in[0,\tilde a]$, let~$z_2(x),\dots,z_q(x)$ denote the
vector corresponding to the asymmetric minimizer of 
$(z_2,\dots,z_q)\mapsto\Phi_{\beta(1-x),0}^{(q-1)}(z_2,\dots,z_q)$ with $z_2=\dots=z_{q-1}<z_q$. Let~$\psi(x)$ denote the quantity~$\Phi_{\beta,h}^{(q)}(\bm)$ evaluated at
$\bm=\bm(x)$ where
\begin{equation}
%\label{5.}
\bm(x)=x\hatv_1+(1-x)z_2(x)\hatv_2+\dots+(1-x)z_q(x)\hatv_q.
\end{equation}
Then
\begin{equation}
\label{eq5.25}
\psi'''(x)<0 \quad \text{for all$~x\in[0,\tilde a]$}.
\end{equation}
\end{lemma}

\begin{proofsect}{Proof}
Let $\psi(x)$ be as stated above. Let $t=t(x)=\beta(1-x)$ and let $\bz(x)=(z_2(x), \dots, z_q(x))$ denote the asymmetric global minimum of $\Phi_{t(x),0}^{(q-1)}$.  This allows us to rewrite $\psi(x)$ as
\begin{multline}
\label{5.1} 
\quad
\psi(x)= -\frac{\beta}{2}x^2 + x\log(x) + (1-x)\log(1-x) -hx
\\+ (1-x)\Phi_{t(x),0}^{(q-1)}(\bz(x)).
\quad
\end{multline}
We will write~$z_2=\dots=z_{q-1}=\frac1{q-1}-\frac{m(t)}{q-2}$ 
and~$z_q=\frac1{q-1} + m(t)$, where $m(t)$ is the maximal positive solution to
\begin{equation}
\label{5.3} 
\frac{q-1}{q-2}m(t)=
\frac{\exp\bigl\{t{\frac{q-1}{q-2}m(t)}\bigr\}-1}
{\exp\bigl\{t\frac{q-1}{q-2}m(t)\bigr\} + q-2}.
\end{equation} 
The various steps of the proof involve two specific functions~$u(t)$ and~$\alpha(t)$ defined by
\begin{equation}
\label{5.4}
u(t)=t\frac{q-1}{q-2}m(t)
\end{equation}
and
\begin{equation} 
\label{5.6}
\alpha(t)=\frac{\texte^{u(t)}}{\texte^{u(t)}+ q-2}
\biggl(1 -\frac{\texte^{u(t)} - 1}{\texte^{u(t)} + q-2}\biggr).
\end{equation}
We state these definitions here to facilitate later reference.

A simple argument gives that $t\mapsto m(t)$ is smooth when $t\geq\betaMF^{(q-1)}$, so $\psi(x)$ is differentiable. The actual proof then commences by the calculation of the third derivative of $\psi(x)$:
\begin{multline}
\quad
\psi'''(x) = -\frac1{x^2} + \frac1{(1-x)^2} 
\\
+ 2\,\frac{q-2}{q-1}\,\Bigl(\frac{u(t)}{1-x}\Bigr)^2
\biggl(3\frac{m'(t)}{m(t)}
+t \frac{m''(t)}{m(t)} + t
\Bigl(\frac{m'(t)}{m(t)}\Bigr)^2
\biggr),
\quad
\end{multline}
where $m'$ and~$m''$ denote the first and second derivative of~$t\mapsto m(t)$ 
and where we have used Lemma~\ref{L1} to differentiate $\Phi_{t,0}^{(q-1)}$.
Since we want to show $\psi'''(x)<0$ and we know that $x\le\tilde a<\ffrac12$, it
suffices to prove the inequality
\begin{equation}
\label{5.5} 
3\frac{m'(t)}{m(t)}
+t \frac{m''(t)}{m(t)} + t
\Bigl(\frac{m'(t)}{m(t)}\Bigr)^2< 0.
\end{equation} 
Differentiating both sides of Eq.~\eqref{5.3} and solving for
$m'(t)$ yields
\begin{equation}
\label{5.7} 
\frac{m'(t)}{m(t)} =
\frac{\alpha(t)}{1 - t \alpha(t)}.
\end{equation}
Taking another derivative with respect to~$t$ allows us to express~$m''(t)/m(t)$ in terms of~$\alpha(t)$ and $\alpha'(t)$. In conjunction with Eq.~\eqref{5.7}, this shows that
Eq.~\eqref{5.5} is equivalent to
\begin{equation}
\label{5.9} 
3+ t \frac{\alpha'(t)}{\alpha(t)}<0.
\end{equation}
Differentiating Eq.~\eqref{5.6} and applying Eqs.~\eqref{5.4} and~\eqref{5.7}, we have
\begin{equation}
\label{5.10} 
\alpha'(t)=\alpha(t)\frac{u(t)}{t[1-t\alpha(t)]}
\biggl(1-2\frac{\texte^{u(t)}}{\texte^{u(t)} + q-2}\biggr).
\end{equation}
Writing Eq.~\eqref{5.9} back in terms of~$u(t)$, we see that Eq.~\eqref{5.5} is equivalent to the inequality
\begin{equation}
\label{5.13} 
3\biggl(1-\frac{t (q-1) \texte^{u(t)}}{(\texte^{u(t)} + q-2)^2}\biggr)
< u(t)\frac{\texte^{u(t)}- q+2}{\texte^{u(t)}+ q-2}.
\end{equation}
The rest of the proof is spent on proving Eq.~\eqref{5.13}.

We first use that~$x\le\tilde a$ implies~$t\ge \betaMF^{(q-1)}=2\frac{q-2}{q-3}\log(q-2)$ and so the left-hand side of Eq.~\eqref{5.13} increases if we replace~$t$ by~$\betaMF^{(q-1)}$. After this, there is no explicit dependence on~$t$ and so we may regard the result as an inequality for the quantity~$u$. Clearing denominators, substituting~$s=\texte^u$, and recalling that~$u(t)\ge2\log(q-2)$ for~$x\le\tilde a$, it suffices to show that
\begin{equation}
\label{5.15} 
\gamma(s)=A_q s+ s^2\log s-\lambda^2\log s-3s^2-3\lambda^2
\end{equation}
is strictly positive for all~$s\ge \lambda^2$ and all~$q\ge4$, where $\lambda=q-2$ and
\begin{equation}
%\label{}
A_q=3(q-1)\betaMF^{(q-1)}-6(q-2).
\end{equation}
Since $\betaMF^{(q-1)}\ge2.5$ for~$q\ge4$, we easily check that~$A_q\ge10$ once~$q\ge4$.

First we will observe that~$\gamma$ is actually increasing for all~$s\ge\lambda^2$. Indeed, a simple calculation shows that, for such~$s$, we have~$\gamma'(s)\ge\omega(s)$, where
\begin{equation}
%\label{}
\omega(s)=A_q-1+2s\log s-5s.
\end{equation}
Next we find that~$\min_{s\ge0}\omega(s)=A_q-1-2\texte^{\ffrac32}$. Since $\texte^{\ffrac32}\approx 4.48$ and $A_q\ge10$, we have that~$\omega$---and hence~$\gamma'$---are strictly positive for~$s\ge\lambda^2$. Hence~$\gamma$ is increasing for all~$s$ of interest.

Once we know that~$\gamma$ is increasing, it suffices to show that~$\gamma(\lambda^2)$ is positive. Here we note that
\begin{equation}
%\label{}
\gamma(\lambda^2)=\frac{q-1}{q-3}(q-2)^2\bigl\{(q^2-3q+6)2\log(q-2)-3(q-1)(q-3)\bigr\}
\end{equation}
and so~$\gamma(\lambda^2)$ is positive once
\begin{equation}
%\label{}
2\log(q-2)>3\,\frac{(q-1)(q-3)}{q^2-3q+6}.
\end{equation}
Noting that the right-hand side is less than~$3$, and using that~$2\log5>3$, this holds trivially for~$q\ge7$. In the remaining cases~$q=4,5,6$, the inequality is verified by direct calculation.
\end{proofsect}

Using Lemma~\ref{lemma5.6} we arrive at the following
conclusion:

\begin{corollary}
\label{cor5.8}
Let $q \geq 4$, $\beta\ge0$ and $h<0$.  Then $\Phi_{\beta,h}^{(q)}$ has at most one (symmetric) global minimizer with~$x_1< x_2=\dots=x_q$ and at most one (asymmetric) global minimizer with~$x_1< x_2=\dots=x_{q-1}<x_q$. 
\end{corollary}

\begin{proofsect}{Proof}
Let~$(x_1,\dots,x_q)$ correspond to a minimizer of~$\Phi_{\beta,h}^{(q)}$. Since~$h<0$, Lemma~\ref{L3} allows us to assume that~$x_1<x_2=\dots=x_{q-1}\le x_q$. If~$x_2=\dots=x_q$, then a simple calculation shows that the quantity~$\theta$, which is related to~$x_1$ via~$x_1=\frac1q-\frac{q-1}q\theta$, obeys the equation~$g(\theta)=\theta$, where~$g$ is as in Eq.~\eqref{eq5.23}. By Lemma~\ref{lemma5.5}, such a solution is unique and so there is at most one symmetric minimizer.

Next let us assume that~$x_q$ exceeds the remaining components. Note that we must have that~$\beta(1-x_1)\ge\betaMF^{(q-1)}$ because otherwise Eq.~\eqref{E3.1} implies that $(x_2,\dots,x_q)$, properly scaled, would correspond to the $(q-1)$-state Potts model in the high-temperature regime. Since in addition~$x_1<\ffrac1q$, we are permitted to use Lemma~\ref{lemma5.6} and conclude that~$x_1$ is a minimizer of the function~$\psi$ from Eq.~\eqref{5.1}. As is seen from its definition and Eq.~\eqref{eq5.25}, $\psi$ starts off convex (and decreasing) at~$x=0$ and, as~$x$ increases, may eventually turn concave. In particular, there could be at most two points in~$[0,\tilde a]$ where~$\psi$ achieves its absolute minimum---one in~$(0,\tilde a)$ and the other at~$\tilde a$.

We claim that if~$\psi'(\tilde a)<0$ then~$\tilde a$ cannot be the first coordinate of an asymmetric global minimizer.
Indeed, if~$\psi$ is strictly decreasing at~$\tilde a$, then the free energy could be lowered by increasing the first component beyond~$\tilde a$. Therefore, if~$\psi'(\tilde a)<0$, then $\psi$ has at most one \emph{relevant} minimum in~$[0,\tilde a]$. On the other hand, the above concavity-convexity picture implies that, once $\psi'(\tilde a)\ge0$, there is \emph{only one} point in~$[0,\tilde a]$ where~$\psi$ is minimized. Hence, in all cases, there is at most one asymmetric minimizer.
\end{proofsect}

The proof of Theorem~\ref{T2} will require some comparisons between the two minimizers allowed by Corollary~\ref{cor5.8}. These are stated in the following~lemma.

\begin{lemma}
\label{lemma-ener}
Let~$q\ge4$,~$\beta\ge0$ and~$h\in(-\infty,0)$. Suppose that~$\Phi_{\beta,h}^{(q)}$ has two minimizers, one symmetric with~$\xS_1<\xS_2=\dots=\xS_q$ and the other asymmetric with~$\xA_1<\xA_2=\dots=\xA_{q-1}<\xA_q$. Then
\begin{equation}
\label{eq5.45}
\xA_1<\xS_1\quad\text{and}\quad\xS_q<\xA_q.
\end{equation}
Moreover, let~$\eA=[\xA_1]^2+\dots+[\xA_q]^2$ and $\eS=[\xS_1]^2+\dots+[\xS_q]^2$. Then there exists a constant~$c_q>0$ such that for any~$h\in[-\infty,0)$ and any~$\beta\ge0$ where both minimizes of~$\Phi_{\beta,h}^{(q)}$ ``coexist,'' we have
\begin{equation}
\label{eq5.46}
\eA-\eS\ge c_q.
\end{equation}
\end{lemma}

Both parts of this lemma are based on the following fact. Let~$(x_1,\dots,x_q)$ be a minimizer of~$\Phi_{\beta,h}^{(q)}$ ordered such that~$x_1<x_2=\dots=x_{q-1}\le x_q$. The stationarity condition yields
\begin{equation}
\label{eq5.47}
x_1\texte^{-\beta x_1-h}=x_2\texte^{-\beta x_2}=\dots=x_q\texte^{-\beta x_q},
\end{equation}
and so let~$\Theta$ denote the common value of this equality. Then we have:

\begin{lemma}
\label{lemma-Theta}
Let~$h<0$ and~$\beta\ge0$. If~$\Theta$ and~$\Theta'$ correspond to two minimizers of~$\Phi_{\beta,h}^{(q)}$, and~$\Theta=\Theta'$, then the minimizers are the same (up to permutations in the last~$q-1$ indices).
\end{lemma}

\begin{proofsect}{Proof}
Suppose that both minimizers are ordered increasingly. By~$h<0$ and Lemma~\ref{L2},~$x_1\le x_2=\dots=x_{q-1}$. The fact that~$(\frac{x_2}{1-x_1},\dots,\frac{x_q}{1-x_1})$ is the minimizer of~$\Phi_{\beta(1-x_1),h}^{(q-1)}$---see Eq.~\eqref{E3.1}---then implies~$\beta x_k\le1$ for all~$k=1,\dots,q-1$. Since the function~$r(x)=x\texte^{-\beta x}$ is invertible for~$x$ with~$\beta x\le1$, equality of the~$\Theta$'s implies equality of the first~$q-1$ coordinates. The constraint on the total sum implies equality of the~$x_q$'s as well.
\end{proofsect}

\begin{proofsect}{Proof of Lemma~\ref{lemma-ener}}
We will first attend to the proof of Eq.~\eqref{eq5.45}. In light of Eq.~\eqref{E3.1}, the $(q-1)$-state Potts system on~$(x_2,\dots,x_q)$ is at the effective temperature~$\betaeffS=(1-\xS_1)\beta$ for the symmetric minimizer and~$\betaeffA=(1-\xA_1)\beta$ for the asymmetric minimizer. But for both symmetric and asymmetric minimizers to ``coexist'' we must have~$\betaeffS\le\beta_\MF^{(q-1)}\le\betaeffA$ and so~$\xA_1\le\xS_1$. To rule out the equality sign, we note that if~$\xA_1=\xS_1$, then the corresponding~$\Theta$'s are the same and Lemma~\ref{lemma-Theta} thus forces equality of all components. Once~$\betaeffS\le\betaeffA$ is known,~$\xS_q<\xA_q$ follows.

In order to prove Eq.~\eqref{eq5.46}, let~$\phi$ be the common value of~$\Phi_{\beta,h}^{(q)}$ for the two minimizers and let~$\ThetaA$ and~$\ThetaS$ be the corresponding~$\Theta$'s. Let us take the logarithm of every term in \eqref{eq5.47}, multiply the result for the~$j$-th term by~$x_j$ and add these all up to get
\begin{equation}
\phi-\frac\beta2\eS=\log\ThetaS
\quad\text{and}\quad
\phi-\frac\beta2\eA=\log\ThetaA.
\end{equation}
As~$\xA_1<\xS_1\le\ffrac1\beta$, we have~$\ThetaA<\ThetaS$ for all~$h\in(-\infty,0)$; for~$h=0,-\infty$ this holds by a direct argument for the zero-field Potts model. Hence~$\eS<\eA$ whenever the two minimizers are ``coexist.'' 

To see that the positivity of $\eA-\eS$ holds uniformly in~$(h,\beta)\in[-\infty,0]\times[0,\infty]$, we use a compactness argument. First, we only need to worry about the~$\beta$'s in a finite, closed interval~$I_q$. Indeed, the effective temperature of the Potts model,~$\betaeff=\beta(1-x_1)$, is a number between~$\beta$ and~$\beta(1-\ffrac1q)$ and so if either~$\beta<\betaMF^{(q-1)}$ or~$\beta(1-\ffrac1q)>\betaMF^{(q-1)}$, then no coexistence of minimizers is possible. 

Next let us consider a sequence of~$(h,\beta)$ in~$[-\infty,0]\times I_q$ with a topology that makes this set compact. If~$\eA-\eS$ tends to zero along this sequence, the above arguments imply that the asymmetric and symmetric minimizers must coalesce as the parameters tend to a limiting point. But this is impossible because by the second half of Eq.~\eqref{eq1}, the scalar magnetization of the corresponding~$(q-1)$-state Potts model, which is proportional to the ratio of $\xA_q-\xA_2$ and~$1-\xA_1$, is always at least~$\frac{q-3}{q-1}$.
\end{proofsect}

\begin{remark}
The previous proof kept the distinctness of~$\eS$ and~$\eA$ in the realm of the existential. A calculation actually shows that, for any~$h<0$, there are constants~$e_1<e_2$ depending only on~$q$ such that~$\eS<e_1$ and~$\eA>e_2$ whenever the two minimizers ``coexist.''
\end{remark}

\begin{proofsect}{Proof of Theorem \ref{TH2}}
Fix~$\beta\ge0$ and~$h<0$. Corollary~\ref{cor5.8} implies that, up to a permutation in all-but-the-first component,~$\Phi_{\beta,h}^{(q)}$ has at most two global minimizers: one symmetric~$\bmS$ and one asymetric~$\bmA$. This proves part~(1) of the theorem.

Among the global minima, the first barycentric coordinate~$x_1=x_1(\beta,h)$ is (strictly) increasing in~$h$ (see Lemma~\ref{L2}) and so the effective coupling $\betaeff(h)=\beta(1-x_1(\beta,h))$, which governs the $(q-1)$-state Potts
model on~$(x_2,\dots,x_q)$, is decreasing. Now if~$\betaeff(h)>\betaMF^{(q-1)}$ then only the asymmetric minimum is relevant, while if~$\beta(h)<\betaMF^{(q-1)}$ then only the symmetric minimum applies. Hence, for~$\beta\in(\betaMF^{(q-1)},\betaMF^{(q)})$, there is a unique~$h_2=h_2(\beta)$ such that the role of minimizers changes as~$h$ increases through~$h_2$. (For~$\beta$ outside $(\betaMF^{(q-1)},\betaMF^{(q)})$, the minimizers are in qualitative agreement with those of~$h=-\infty$ or $h=0^-$.) In particular, the minimizer is unique for~$h\ne h_2(\beta)$ and both minimizers ``coexist'' for~$h=h_2(\beta)$. 

Modulo the definition of function~$\beta_-^{(q)}$, parts~(2-4) of the theorem are proved. It remains to show that $\beta\mapsto h_2(\beta)$ is strictly increasing (and thus invertible), continuous and with limits~$-\infty$ and~$0$ at the left and right endpoints of $(\betaMF^{(q-1)},\betaMF^{(q)})$, respectively. By Lemma~\ref{lemma-ener}, the quantities~$\eS$ and~$\eA$ are separated by a ``gap.'' A simple limiting argument (not dissimilar to that used in the proof of Theorem~\ref{TH1}) now shows that~$h_2$ is continuous. Moreover, by Lemma~\ref{L1}, the norm-squared of all minimizers increases with~$\beta$, and so~$h_2$ is strictly monotone and the limits of~$h_2$ at the endpoints of $(\betaMF^{(q-1)},\betaMF^{(q)})$ must be as stated. These facts allow us to define~$\beta_-^{(q)}$ as the inverse of~$h_2$ and verify all its properties in part~(5) of the theorem. 
\end{proofsect}

\section{Proofs: Actual systems}
\label{sec6}\noindent 
Here we will provide the proofs of our results for actual spin systems. The main portion of the arguments has already been given in Sects.~\ref{sec4} and~\ref{sec5}. We will draw freely on the notation from these sections. The proofs are fairly straightforward (and mostly existential) and so we will stay rather brief.

\smallskip
First we will attend to the zero-field Potts model:

\begin{proofsect}{Proof of Theorem \ref{T1}}
The proof is more or less identical to that of Theorem~2.1 of Ref.~\cite{BC}; the only substantial difference is that now we are not permitted to assume that the magnetization is monotone (indeed, some of the $J_{x,y}$'s may be negative). We will base our arguments on the mean-field properties of the zero-field Potts model, as outlined in Sect.~\ref{sec2.2}.

Recall the mean-field free-energy function~$\Phi_{\beta,0}^{(q)}$ from Eq.~\eqref{1.15a}. By the fact that the global minimizer of~$\Phi_{\beta,0}^{(q)}$ changes from symmetric to asymmetric as~$\beta$ increases through~$\betaMF^{(q)}$, we can make the following conclusions: Given~$\beta\approx\betaMF^{(q)}$, let~$\UU_\epsilon$ be an~$\epsilon$-neighborhood of~$\bm=\boldsymbol0$ and let~$\VV_\epsilon$ be the union of $\epsilon$-neighborhoods of the asymmetric minimizers. Then for each~$\epsilon>0$, there exists~$\delta>0$ such that for all~$\beta$ with~$|\beta-\betaMF^{(q)}|\le\epsilon$ the set
\begin{equation}
\label{6.1}
\OO_\delta=\bigl\{\bm\in\Conv(\Omega)\colon\Phi_{\beta,0}^{(q)}(\bm)-F_\MF(\beta,0)<\delta\bigr\}
\end{equation}
is contained in $\UU_\epsilon\cup\VV_\epsilon$. Moreover, if~$\beta=\betaMF^{(q)}-\epsilon$, then $\OO_\delta\subset\UU_\epsilon$ while at~$\beta=\betaMF^{(q)}+\epsilon$, we have $\OO_\delta\subset\VV_\epsilon$. 

Let~$\scrM_\star(\beta,0)$ be the set of ``extremal magnetizations.'' By Theorem~\ref{thm2.2}, if the integral~$\scrI$ in Eq.~\eqref{2.5} is so small that~$\beta\frac\kappa2n\scrI=\beta\frac{q-1}2\scrI\le\delta$ for all~$\beta$ with $\beta\le\betaMF^{(q)}+\epsilon$, then~$\scrM_\star\subset\OO_\delta$. Now the asymmetric minimizers have norm at least~$\ffrac12$, and the near-monotonicity of the magnetization from Lemma~\ref{lemma4.8} thus implies that, at some~$\betat$ with~$|\betat-\betaMF^{(q)}|\le\epsilon$, the physical magnetization jumps from some value inside~$\UU_\epsilon$ to some value inside~$\VV_\epsilon$. The jump (of this size) is unique by Lemma~\ref{lemma4.8}. From here the claims \twoeqref{eq3.17}{eq3.19} follow.
\end{proofsect}

Next we dismiss the cases with non-zero field:

\begin{proofsect}{Proof of Theorem \ref{T2}}
Let~$\hc$ be the quantity from Theorem~\ref{TH1} and~$\betaMF^{(q)}(h)$ be the concatenation of functions~$\beta_+$ and~$\beta_-$ from Theorems~\ref{TH1} and~\ref{TH2}. An argument similar to the one used in the previous proof shows that, for each~$\epsilon>0$ there exists~$\delta>0$, such that if~$\betaMF^{(q)}\frac\kappa2n\scrI\le\delta$ and~$h\le\hc-\epsilon$, a strong first-order transition occurs at some~$\betat(h)$ which is within~$\epsilon$ of~$\betaMF^{(q)}(h)$. This transition is manifested by a jump in both magnetization and energy density. This proves part~(1) of the theorem.

As to part~(2), by Lemma~\ref{lemma-ener} we know that the first components of the two minimizers are uniformly separated whenever~$h$ is confined to a compact subset of~$(-\infty,\hc)$. Since our general bounds in Theorem~\ref{thm2.2} imply that the physical magnetizations at~$(h,\betat(h))$ are very near their mean-field values provided~$\scrI$ is sufficiently small, also the first components thereof must be different. Using the monotonicity of the first component of physical minimizers in~$h$, the existence of a jump in~$m_\star(\beta,h)$ on the transition line follows.
\end{proofsect}

\begin{proofsect}{Proof of Theorem~\ref{BCT2}}
The proof is based on Theorem~\ref{BCT1} and Lemma~\ref{lemma-BCmax}. Indeed, Theorem~\ref{BCT1} implies that all minima are characterized by the fact that one of~$(x_1,x_0,x_{-1})$ is larger than~$1-C\texte^{-\beta}$. These minima are nearly degenerate for~$\lambda$ of order~$\texte^{-\beta}$ with free energy difference given by~$\lambda-\texte^{-\beta}+O(\beta\texte^{-\beta})$. The goal is to show that the free energy is uniformly large (on the scale of~$\texte^{-\beta}$ in the complement of the~$C\texte^{-\beta}$-neighborhood of these minima.

Let~$C\gg1$ be the number exceeding the corresponding constant from Theorem~\ref{BCT1} and suppose that $|\lambda|\le C\texte^{-\beta}$. Consider the set~$\OO_\beta$ of all triplets~$(x_1,x_0,x_{-1})$ with~$x_1+x_0+x_{-1}=1$, such that~$\max\{x_1,x_0,x_{-1}\}>1-C\texte^{-\beta}$. We claim that for~$\beta\ge\beta_0$ (with~$\beta_0$ depending on~$C$),
\begin{equation}
\label{eq6.2}
\inf_{(x_1,x_0,x_{-1})\in\OO_\beta^\cc}\Phi_{\beta,\lambda}(x_1,x_0,x_{-1})\ge\alpha(C\log C)\texte^{-\beta},
\end{equation}
where~$\alpha$ is a positive number independent of~$C$. Indeed, Theorem~\ref{BCT1} implies that all local minima of~$\Phi_{\beta,\lambda}$ lie in~$\OO_\beta$, and so the absolute minimum of~$\Phi_{\beta,\lambda}$ must occur on the boundary of~$\OO_\beta^\cc$. But the ``outer'' boundary of~$\OO_\beta^\cc$ is not a possibility, and so the mimimum occurs at a point with~$\max\{x_1,x_0,x_{-1}\}=1-C\texte^{-\beta}$. The bound \eqref{eq6.2} is then a consequence of Lemma~\ref{lemma-BCmax}.

Let now the integral~$\scrI$ in Eq.~\eqref{2.5} be such that~$\beta\scrI\ll(C\log C)\texte^{-\beta}$. Then Theorem~\ref{thm2.2} ensures that all physical magnetizations (from~$\scrM_\star$) are contained inside~$\OO_\beta$. However, by Eq.~\eqref{3.16b}, for~$\beta$ such that~$\lambda-\texte^{-\beta}\ge O(\beta\texte^{-\beta})$ the set~$\OO_\beta$ contains no triplets with dominant~$x_{\pm1}$ while for $\lambda-\texte^{-\beta}\le O(\beta\texte^{-\beta})$, there are no~$x_0$-dominant states. The standard thermodynamic arguments imply that the amount of zero-ness decreases as~$\lambda$ increases. Hence, there must be a jump at some~$\lambdat=\texte^{-\beta}+O(\beta\texte^{-\beta})$ from states dominated by~$0$'s to those where~$0$'s are very sparse. This finishes the proof.
\end{proofsect}

\section*{Acknowledgments}
\noindent 
This research was supported by the NSF grant~DMS-0306167.

\end{document}